\documentclass[lettersize,journal]{IEEEtran}
\usepackage{amsmath,amsfonts,amssymb}
\usepackage{algorithmic}
\usepackage{algorithm}
\usepackage{array}
\usepackage[caption=false,font=normalsize,labelfont=sf,textfont=sf]{subfig}
\usepackage{textcomp}
\usepackage{stfloats}
\usepackage{url}
\usepackage{verbatim}
\usepackage{graphicx}
\usepackage{cite}
\usepackage{booktabs}
\usepackage{multirow}
\usepackage{makecell}
\usepackage{balance}
\usepackage[colorlinks=true, linkcolor=blue, citecolor=blue]{hyperref}
\usepackage[margin=0.75in]{geometry}
\usepackage{tikz}

\newcommand{\llbracket}{[\![}
\newcommand{\rrbracket}{]\!]}
\usepackage{enumitem}
\usepackage{caption}

\newsavebox{\protobox}
 
\newcommand{\dashedline}{%
  \par\vspace{4pt}%
  \tikz\draw[dashed, dash pattern=on 3pt off 2pt] (0,0) -- (\linewidth,0);%
  \par\vspace{4pt}%
}

\begin{document}

\title{Efficient and High-Accuracy Private CNN Inference \\with Helper-Assisted Malicious Security}

\author{Kaiwen~Wang,
        Xiaolin~Chang,
        Junchao~Fan,
        and~Yuehan~Dong
\thanks{K. Wang, X. Chang, J. Fan, and Y. Dong are with the Beijing Key Laboratory of Security and Privacy in Intelligent Transportation, Beijing Jiaotong University, China.}%
}

\markboth{IEEE Transactions on XXXXX,~Vol.~XX, No.~X, Month~2025}%
{Wang \MakeLowercase{\textit{et al.}}: Efficient Private Inference Based on HA-MSDM MPC}

\maketitle

\begin{abstract}
Machine Learning as a Service (MLaaS) exposes sensitive client data to service providers. Private inference mitigates this risk while preserving model functionality. Despite extensive progress in MPC-based solutions, they remain constrained by a fundamental three-way tension among strong security, efficiency, and model accuracy. This challenge is particularly acute under the malicious dishonest majority (MSDM) setting, where prior work either incurs high communication overhead or suffers non-negligible accuracy loss due to polynomial approximations of nonlinear functions. Although the helper-assisted MSDM (HA-MSDM) model improves efficiency and fairness, it lacks a dedicated design for accurate and efficient neural network inference. In this work, we present an HA-MSDM-based private CNN inference framework that simultaneously achieves high efficiency and near-plaintext accuracy through a co-design of cryptographic primitives, MPC protocols, and model training. Specifically, we (i) extend authenticated sharing to rings to enable efficient fixed-point computation, (ii) design constant-round protocols for multiplication and polynomial evaluation, with round complexity independent of the polynomial degree, and (iii) introduce a training strategy that recovers the expressiveness of polynomial models via knowledge distillation and warm initialization. Experiments demonstrate 2.3--6.8$\times$ speedup in LAN and 1.3--5.6$\times$ in WAN over state-of-the-art MSDM frameworks, while achieving accuracy within 0.5\% of ReLU-based plaintext models.
\end{abstract}


\section{Introduction}

\IEEEPARstart{M}{achine} Learning as a Service (MLaaS) platforms offered by major cloud providers such as Amazon, Microsoft, and Google~\cite{ref1} have greatly lowered the barrier to deploying machine learning models. However, MLaaS requires clients to submit sensitive data to the service provider, exposing them to serious privacy risks. Such data may include medical records, financial transactions, or facial images. Private inference~\cite{ng2023sok} addresses this problem. It enables a client to obtain the model's prediction on their input without revealing the input to the service provider, and without the client learning the model parameters.

Among existing private inference approaches, those based on Secure Multi-party Computation (MPC)~\cite{yao1982protocols} have emerged as a leading direction. MPC provides provable security under well-defined cryptographic assumptions. It allows mutually distrusting parties to jointly compute a function over their private inputs while revealing only the final output. The security of an MPC protocol is characterized by two dimensions. The first dimension is the adversary's capability: a \textit{semi-honest} adversary follows the protocol but attempts to infer secrets, while a \textit{malicious} adversary may deviate arbitrarily. The second dimension is the number of corrupted parties: \textit{honest majority} allows less than half of the parties to be corrupted, while \textit{dishonest majority} allows up to $n-1$ of $n$ parties to be corrupted.

The Malicious Security Dishonest Majority (MSDM) model~\cite{cramer2018spd,yuan2024md,zhang2025md} captures the most stringent and practically relevant threat setting. This is because malicious adversaries have strong incentives to collude in real deployments. However, MSDM-based frameworks face two drawbacks. First, they suffer from an order-of-magnitude efficiency gap compared to frameworks under weaker threat models. Second, they only guarantee abort security, which allows a malicious adversary to terminate the protocol after learning the output, preventing honest parties from obtaining it.

\begin{figure}[!t]
\centering
\includegraphics[width=3.5in]{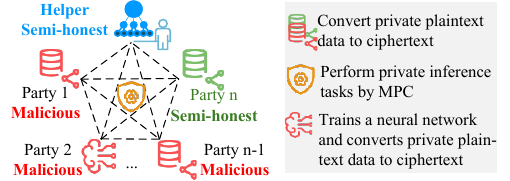}
\caption{Private inference network topology based on the HA-MSDM model.}
\label{fig:topology}
\end{figure}

To address these drawbacks, Asterisk~\cite{karmakar2024asterisk} introduced the \textbf{Helper-Assisted Malicious Security Dishonest Majority (HA-MSDM)} model. The HA-MSDM model augments the MSDM setting with a single semi-honest, non-colluding helper party (HP). The HP can be instantiated as the inference service provider that holds no input data. This additional trust assumption brings two benefits. First, it enables substantially improved protocol efficiency through HP-generated preprocessing material. Second, it achieves stronger \textit{fairness} output security~\cite{cleve1986limits}, ensuring that either all parties receive the output or none do. Fig.~\ref{fig:topology} illustrates the network topology of HA-MSDM-based private inference.

\textbf{Goal of this paper.} We aim to design a private inference framework under the HA-MSDM model that simultaneously achieves \textbf{high efficiency} and \textbf{high accuracy}. High efficiency requires low latency under both LAN and WAN networks. High accuracy requires the ciphertext inference accuracy to closely match the plaintext inference accuracy of the original neural network. Simultaneously achieving both is non-trivial, because techniques that improve efficiency often degrade accuracy, and vice versa.

\textbf{Challenges.} Realizing this goal over HA-MSDM faces three intertwined challenges.

\textbf{Challenge 1 (Primitive efficiency):} The only existing HA-MSDM framework, Asterisk, is a general-purpose MPC framework built over the prime field $\mathbb{F}_p$. It supports only basic operations such as addition, multiplication, and dot product. Private inference, however, requires frequent fixed-point arithmetic. Truncation over $\mathbb{F}_p$ requires costly division, while truncation over rings $\mathbb{Z}_{2^l}$ reduces to efficient bit-shift operations~\cite{damgaard2019new}. A direct substitution of $\mathbb{F}_p$ with $\mathbb{Z}_{2^l}$ breaks the security of MAC-based authentication, allowing a malicious adversary to tamper with data with probability $1/2$~\cite{cramer2018spd}. The first challenge is to construct secure sharing semantics over rings that support efficient fixed-point arithmetic in the HA-MSDM model.

\textbf{Challenge 2 (Nonlinear layer efficiency):} Nonlinear layers in CNNs account for the dominant portion of private inference latency. Two mainstream approaches exist. The first uses garbled circuits or binary secret sharing~\cite{yuan2024md,zhang2025md}, which requires $O(\log l)$ communication rounds per nonlinear operation and becomes prohibitively slow in WAN settings. The second uses polynomial approximation~\cite{garimella2021sisyphus,diaa2024fast}, which computes a degree-$k$ polynomial to approximate the target activation function. The naive approach for polynomial evaluation requires $O(k)$ sequential multiplications, each incurring one communication round. Two standard optimizations exist: Horner's method still requires $k$ rounds, while the binary tree method reduces this to $\lceil\log_2 k\rceil$ rounds. Both techniques remain round-dependent on $k$ and incur $O(nk)$ online communication. Existing fixed-round polynomial protocols~\cite{diaa2024fast,lu2019honeybadgermpc} cannot be directly applied to HA-MSDM due to integer overflow or truncation errors. The second challenge is to design fixed-round polynomial evaluation protocols compatible with HA-MSDM sharing semantics.

\textbf{Challenge 3 (Inference accuracy):} Polynomial approximation of activation functions like ReLU only provides high precision within a prescribed fitting interval. Outside this interval, the polynomial output diverges sharply, a phenomenon known as the \textit{escaping activation problem}~\cite{garimella2021sisyphus}. State-of-the-art work PILLAR~\cite{diaa2024fast} addresses this through a regularization term during training that constrains activation inputs to stay within the fitting interval. While effective at constraining activations, this regularization simultaneously limits the model's expressive capacity. It leaves a 2--5\% accuracy gap between polynomial models and the original ReLU baseline---an unacceptable loss for accuracy-sensitive applications. The third challenge is to close this accuracy gap without sacrificing the constraining effect of regularization.

\begin{figure}[!t]
\centering
\includegraphics[width=3.5in]{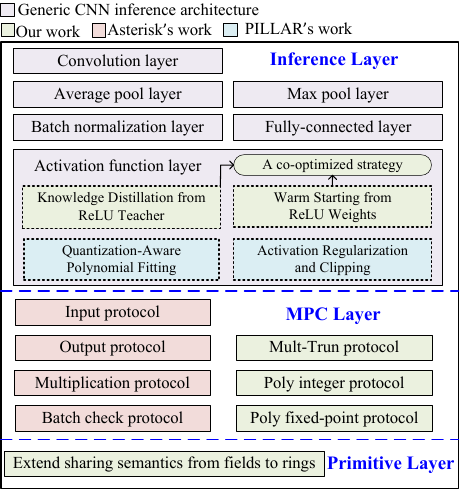}
\caption{The three-layer structure of our framework.}
\label{fig:framework}
\end{figure}

\textbf{Our framework.} To address these challenges, we propose a three-layer private inference framework over HA-MSDM, illustrated in Fig.~\ref{fig:framework}. To the best of our knowledge, this is among the first works to co-design cryptographic primitives, MPC protocols, and training strategies specifically for CNN private inference under the HA-MSDM model, simultaneously targeting high efficiency and near-plaintext accuracy. The three layers and their core contributions are:

\begin{enumerate}

\item \textbf{Primitive layer (addresses Challenge 1):} We design a new authenticated-ring sharing scheme tailored to the HA-MSDM model. The scheme augments Asterisk's masked sharing with a ring-domain MAC construction whose tampering probability is bounded at $2^{-s+\lceil\log(s+1)\rceil}$ for statistical security parameter $s$. Unlike a direct replacement of $\mathbb{F}_p$ with $\mathbb{Z}_{2^l}$ (which would break MAC soundness with constant probability), our scheme separates the computation domain $\mathbb{Z}_{2^{l+s}}$ from the correctness domain $\mathbb{Z}_{2^l}$, and preserves HP-assisted preprocessing efficiency. This is the foundation that enables both efficient fixed-point truncation and the fixed-round polynomial protocols below; neither is achievable over Asterisk's original $\mathbb{F}_p$ domain.

\item \textbf{MPC layer (addresses Challenge 2):} We design three efficient MPC protocols: $\pi_{\text{multTrun}}$ for fixed-point multiplication with truncation, $\pi_{\text{poly\_integer}}$ for integer polynomial evaluation, and $\pi_{\text{poly\_fixed}}$ for fixed-point polynomial evaluation. All three protocols complete within 2 fixed communication rounds independent of the polynomial degree $k$. TABLE~\ref{tab:comparison_theory} summarizes the theoretical communication costs compared with Horner's method and the binary tree method applied to Asterisk.

\item \textbf{Inference layer (addresses Challenge 3):} We design a co-optimized training strategy that combines PILLAR's activation regularization with two novel techniques---knowledge distillation from a ReLU teacher and warm starting from ReLU weights---to recover the expressive capacity lost due to regularization. Our strategy closes the 2--5\% accuracy gap to within 0.5\% across five CNN architectures while preserving the input-constraining effect of regularization.
\end{enumerate}

\begin{table}[!t]
\caption{Comparison of Theoretical Communication Cost and Rounds\label{tab:comparison_theory}}
\centering
\setlength{\tabcolsep}{3pt}
\renewcommand{\arraystretch}{1.15}
\begin{tabular}{l|l|c|c|c}
\hline
\multirow{2}{*}{\textbf{Operation}} & \multirow{2}{*}{\textbf{Framework}} & \multicolumn{2}{c|}{\textbf{Comm. (elem.)}} & \multirow{2}{*}{\makecell{\textbf{Online}\\\textbf{Rounds}}} \\
\cline{3-4}
 & & \textbf{Preprocess} & \textbf{Online} & \\
\hline
\makecell[l]{Mult. with\\Truncation} & \textbf{Ours} ($\pi_{\text{multTrun}}$) & $5$ & $2(n{-}1)$ & $2$ \\
\hline
\multirow{3}{*}{\makecell[l]{Integer\\Poly.}} & Asterisk (Horner) & $3k$ & $2nk$ & $2k$ \\
 & Asterisk (Bin. Tree) & $3(k{-}1)$ & $2n(k{-}1)$ & $2\lceil\log k\rceil$ \\
 & \textbf{Ours} ($\pi_{\text{poly\_integer}}$) & $2k{+}2$ & $2(n{-}1)$ & $2$ \\
\hline
\makecell[l]{Fixed-point\\Poly.} & \textbf{Ours} ($\pi_{\text{poly\_fixed}}$) & $2k{+}4$ & $2(n{-}1)$ & $2$ \\
\hline
\multicolumn{5}{l}{\footnotesize $n$: the number of parties; $k$: polynomial degree.}\\
\multicolumn{5}{l}{\footnotesize Asterisk supports neither multiplication-with-truncation nor fixed-}\\
\multicolumn{5}{l}{\footnotesize point polynomial evaluation, hence not listed.}\\
\end{tabular}
\end{table}

\textbf{Implementation and evaluation.} We implemented our framework in C++20 and conducted extensive evaluations on seven CNN architectures---LeNet, AlexNet, ResNet-18/32/110, VGG-16, and MiniONN---across MNIST, CIFAR-10, CIFAR-100, Tiny-ImageNet, and ImageNet datasets. Our framework achieves 2.3--6.8$\times$ speedup in LAN and 1.3--5.6$\times$ speedup in WAN over state-of-the-art MSDM frameworks MD-ML~\cite{yuan2024md} and MD-SONIC~\cite{zhang2025md}, while matching the ReLU baseline accuracy within 0.5\%.

\textbf{Paper organization.} Section~\ref{sec:prelim} introduces the threat model and preliminaries. Section~\ref{sec:related} reviews related work. Sections~\ref{sec:primitive},~\ref{sec:protocols}, and~\ref{sec:inference} detail the primitive layer, the MPC layer, and the inference layer, respectively. Section~\ref{sec:security} analyzes the security of our protocols. Section~\ref{sec:experiments} presents the experimental evaluation. Section~\ref{sec:conclusion} concludes the paper.

\section{Preliminaries}\label{sec:prelim}

\subsection{Threat Model}\label{subsec:threat}

We consider a distributed computation system comprising $n$ parties $\mathcal{P} = \{P_1,\ldots,P_n\}$, illustrated in Fig.~\ref{fig:topology}. Each party $P_i$ holds private input $x_i \in \{0,1\}^*$ and independent randomness $r_i \in \{0,1\}^*$. The computation proceeds in synchronous rounds over a fully connected network with private and authenticated point-to-point channels. An auxiliary entity HP operates outside $\mathcal{P}$, maintaining a secure channel with each party. HP neither possesses inputs nor receives outputs, but is permitted to be stateful.

We consider a non-colluding adversary that can either maliciously corrupt up to $n-1$ out of $n$ parties, or semi-honestly corrupt HP. The non-colluding property ensures that a semi-honest adversary cannot obtain the view of a malicious adversary. This model, referred to as the Helper-Assisted Malicious Security Dishonest Majority (HA-MSDM) model, was introduced by Asterisk~\cite{karmakar2024asterisk}. Compared to the standard MSDM model~\cite{cramer2018spd,yuan2024md,zhang2025md}, the HA-MSDM model enables stronger output security (fairness rather than abort) and more efficient protocol design by leveraging the semi-honest HP.

\subsection{Asterisk's Sharing Semantics and Structure}\label{subsec:asterisk}

Our framework builds upon the sharing semantics and sub-protocols of Asterisk~\cite{karmakar2024asterisk}, which we summarize below. The main notations are listed in TABLE~\ref{tab:notation}.

\subsubsection{Sharing Semantics}

Asterisk defines three sharing semantics over a finite field $\mathbb{F}_p$ (which we later extend to rings in Section~\ref{sec:primitive}):

\begin{enumerate}
\item \textbf{Additive Share $[x]$:} Each party $P_i$ holds $[x]_i \in \mathbb{F}_p$ such that $\sum_{i=1}^{n}[x]_i = x$.

\item \textbf{MAC Share $\langle x \rangle$:} Each party $P_i$ holds $\langle x \rangle_i = ([x]_i, [t_x]_i)$, where $t_x = \alpha \cdot x$ is the authentication tag under a global MAC key $\alpha$ additively shared among parties.

\item \textbf{Masked Share $\llbracket x \rrbracket$:} Each party $P_i$ holds $\llbracket x \rrbracket_i = (m_x, \langle \lambda_x \rangle_i)$, where $\lambda_x$ is a random mask and $m_x = x + \lambda_x$ is the masked value known to all parties.
\end{enumerate}

All three schemes are linear: given shares of $a_1, \ldots, a_m$ and public constants $c_1, \ldots, c_m$, shares of $\sum_{i} c_i a_i$ can be computed non-interactively.

\subsubsection{Protocol Structure}

Asterisk operates in three phases:

\textbf{Setup phase.} Parties and HP establish shared Pseudo Random Function (PRF) keys: each party $P_i$ shares a key $k_i$ with HP, the parties share a common key $k_{\mathcal{P}}$ (unknown to HP), and a global key $k_{\text{all}}$ is shared among $\mathcal{P} \cup \{\text{HP}\}$. These keys enable non-interactive generation of common random values.

\textbf{Preprocessing phase.} This input-independent phase generates authenticated mask shares for all circuit wires. HP knows the plaintext mask $\lambda_x$ for each wire, while parties hold $\langle \lambda_x \rangle$. Three sub-protocols are used:
\begin{itemize}
\item $\mathcal{F}_{\text{sh}}(\text{Rand}, P_d)$: Generates $\langle \cdot \rangle$-sharing of a random value known to dealer $P_d$ and HP (2 elements).
\item $\mathcal{F}_{\text{sh}}(\text{Rand}, \text{HP})$: Generates $\langle \cdot \rangle$-sharing of a random value known only to HP (1 element).
\item $\mathcal{F}_{\text{sh}}(v, \text{HP})$: Generates $\langle \cdot \rangle$-sharing of a value $v$ held by HP (2 elements).
\end{itemize}
Since only HP sends messages in this phase, the preprocessing is inherently error-free.

\textbf{Online phase.} Once inputs are available, the online phase generates masked values for all circuit wires using $P_{\text{king}}$-based reconstruction~\cite{damgaard2007scalable}: parties send their shares to a designated $P_{\text{king}}$, who reconstructs and broadcasts the result. Before output reconstruction, a batched MAC verification detects any malicious behavior. HP releases output masks only if verification passes, ensuring fairness.

\begin{table}[!t]
\caption{Notation Used Throughout This Paper\label{tab:notation}}
\centering
\begin{tabular}{c|l}
\hline
\textbf{Notation} & \textbf{Description} \\
\hline
$[1,n]$ & The set of integers $\{1, 2, \ldots, n\}$ \\
$P_1, \ldots, P_n$ & $n$ parties in the secure computation \\
HP & The semi-honest helper party \\
$[x]$ & Additive sharing of value $x$ \\
$\langle x \rangle$ & MAC sharing of value $x$ \\
$\llbracket x \rrbracket$ & Masked sharing of value $x$ \\
$\lambda_x$ & The random mask of value $x$ \\
$\alpha$ & The global MAC key \\
$t_x$ & Authentication tag, $t_x = \alpha \cdot x$ \\
$m_x$ & Masked value, $m_x = x + \lambda_x$ \\
$\mathbb{Z}_{2^l}$ & The ring of $l$-bit elements \\
$s$ & The statistical security parameter \\
$d$ & The number of fractional bits in fixed-point \\
$x \equiv_l y$ & $x = y \bmod 2^l$ \\
\hline
\end{tabular}
\end{table}

\section{Related Work}\label{sec:related}

We review related work from two perspectives: MPC models with security guarantees similar to ours, and techniques for handling nonlinear layers in private inference.

\subsection{Malicious Security Dishonest Majority MPC and Variants}
\label{rel:variants}

\subsubsection{MSDM MPC}

SPDZ~\cite{damgaard2012multiparty,damgaard2013practical} secures MPC via MAC-authenticated shares over finite fields. MASCOT~\cite{keller2016mascot} improved SPDZ's offline phase by replacing homomorphic encryption with oblivious transfer, and Overdrive~\cite{keller2018overdrive} further enhanced preprocessing via additive homomorphic encryption. SPDZ$_{2^k}$~\cite{cramer2018spd} extended SPDZ from finite fields to rings $\mathbb{Z}_{2^k}$, enabling native integer arithmetic. Turbospeedz~\cite{ben2019turbospeedz} reduced online communication per multiplication from two openings to one via function-dependent preprocessing. Subsequent works~\cite{orsini2020overdrive2k,escudero2022more} further improved preprocessing efficiency over rings. Despite these advances, MSDM frameworks retain an order-of-magnitude efficiency gap compared to frameworks under weaker threat models.

\subsubsection{MPC with Auxiliary Entities}

To bridge this gap, researchers explored MPC models with auxiliary entities. Early works~\cite{chandran2019universally,hazay2016composable,badrinarayanan2018non} assumed fully honest helpers with minimal capabilities. Ishai et al.~\cite{ishai2022fully} first modeled the helper as semi-honest and non-colluding, achieving fairness but with poor efficiency. Muth et al.~\cite{muth2022assisted} considered probabilistic semi-honest helpers, improving efficiency but achieving only abort security. Asterisk~\cite{karmakar2024asterisk} achieved both efficiency and fairness in the HA-MSDM model (detailed in Section~\ref{subsec:asterisk}). However, extending Asterisk into a private inference framework faces two limitations:

\textbf{1) The inefficient handling on prime field $\mathbb{F}_p$.} Asterisk operates over the prime field $\mathbb{F}_p$, where fixed-point truncation requires costly division rather than efficient bit-shift operations available in rings~\cite{damgaard2019new}. Our work addresses this by extending the sharing semantics from $\mathbb{F}_p$ to $\mathbb{Z}_{2^{l+s}}$ (Section~\ref{sec:primitive}).

\textbf{2) The inefficient MPC protocols required for private inference.} Asterisk provides only basic operations (addition, multiplication, dot product). Its less-than-zero (LTZ) protocol requires $2\lceil\log l\rceil + 4$ communication rounds with $O(n\log_4 l)$ communication complexity, which is impractical for the frequent nonlinear computations in private inference. Our work addresses this by designing fixed-round multiplication-truncation and polynomial evaluation protocols (Section~\ref{sec:protocols}).

\subsection{Nonlinear Layers Processing in Private Inference}

\textbf{Protocol Conversion.} MD-ML~\cite{yuan2024md} extends circuit-dependent preprocessing to SPDZ$_{2^k}$ and employs mask-and-compare methods for secure comparison, enabling ReLU computation under the MSDM model. MD-SONIC~\cite{zhang2025md} represents the current state-of-the-art in MSDM private inference by combining SPDZ$_{2^k}$-sharing for arithmetic circuits with TinyOT-sharing~\cite{burra2021high} for Boolean circuits, using edaBits and daBits for efficient mixed-circuit protocol conversion.

\textbf{Polynomial Approximation.} Early works~\cite{gilad2016cryptonets,chabanne2017privacy,hesamifard2018privacy} demonstrated polynomial-based private inference on shallow networks with limited accuracy. Delphi~\cite{mishra2020delphi} proposed a hybrid approach selecting between garbled circuits and polynomial computation per ReLU, but incurs substantial GC overhead. Lee et al.~\cite{lee2023precise} achieved precise approximation using 29th-degree polynomials at prohibitive computational cost. Sisyphus~\cite{garimella2021sisyphus} first systematically identified the escaping activation problem and proposed approximate min-max normalization, but suffers significant accuracy degradation in deep CNNs(more than 11-layer). PILLAR~\cite{diaa2024fast} introduced activation regularization and clipping, substantially improving the accuracy of polynomial-based inference in deep CNNs. However, PILLAR's regularization simultaneously constrains the model's expressiveness, resulting in a 2--5\% accuracy gap between polynomial and ReLU models. Our work (Section~\ref{sec:inference}) closes this gap through knowledge distillation and warm starting while preserving PILLAR's effectiveness in constraining activation inputs. TABLE~\ref{tab:comparison_framework} presents a comparison of our framework with state-of-the-art private inference frameworks.

\textbf{MPC Protocols for Polynomial Evaluation.} Several lines of work have studied constant-round polynomial evaluation in different MPC models, but none directly applies to HA-MSDM. Damg{\aa}rd et al.~\cite{damgaard2006unconditionally} and Polymath~\cite{lu2019honeybadgermpc} achieve 2-round evaluation via preprocessed $[r^{-n}]$ and power-of-$r$ shares, but require honest-majority Shamir sharing. ABY2.0~\cite{patra2021aby2} achieves 1-round $N$-input multiplication but targets 2PC semi-honest security with communication linear in fan-in, without truncation over rings. PILLAR's ESPN~\cite{diaa2024fast} evaluates $(x_A + x_B)^k$ via binomial expansion in a single round, but this construction is specific to 2PC semi-honest additive sharing: lifting it to HA-MSDM requires MAC-authenticated sharing of every $O(k)$ cross-term $x_A^i \cdot x_B^{k-i}$, which inflates both preprocessing cost and ring width, and generalizing to $n>2$ parties turns the binomial expansion into a multinomial one with $O(k^{n-1})$ cross-terms. Our $\pi_{\text{poly\_fixed}}$ instead computes all powers via a local DP recurrence on HP-preprocessed $\llbracket r^j \rrbracket$, yielding $2$ rounds and $2(n-1)$ elements online, independent of both $k$ and fan-in, while supporting fixed-point truncation and malicious security.

\begin{table*}[!t]
\caption{Comparison with State-of-the-art Private Inference Frameworks\label{tab:comparison_framework}}
\centering
\renewcommand{\arraystretch}{1.15}
\begin{tabular}{l|c|c|c|c|c|c}
\hline
\textbf{Framework} & \textbf{Parties} & \textbf{Threat Model} & \textbf{Output Security} & \textbf{Nonlinear Method} & \textbf{Activation Support} & \textbf{110-layer High Prec.$^\S$} \\
\hline
Sisyphus~\cite{garimella2021sisyphus} & 3+ & Unspecified$^\dagger$ & abort & Poly (ASS) & Unlimited & $\times$ \\
PILLAR~\cite{diaa2024fast} & 2 & Unspecified$^\dagger$ & abort & Poly (ASS) & Unlimited & $\times$ \\
MD-ML~\cite{yuan2024md} & 3+ & MSDM & abort & BSS$^\ddagger$ & ReLU & $\checkmark$ \\
MD-SONIC~\cite{zhang2025md} & 3+ & MSDM & abort & BSS$^\ddagger$ & ReLU & $\checkmark$ \\
\textbf{Ours} & 2+ & HA-MSDM & fairness & Poly (ASS) & Unlimited & $\checkmark$ \\
\hline
\multicolumn{7}{l}{\footnotesize $\dagger$: No specific MPC model is designated; techniques are transferable to general MPC models.} \\
\multicolumn{7}{l}{\footnotesize $\ddagger$: BSS-based nonlinear computation is typically 1--2 orders of magnitude slower than ASS-based.} \\
\multicolumn{7}{l}{\footnotesize $\S$: High Precision: cipher accuracy $\leq$1\% below plaintext accuracy with precise activation functions.} \\
\end{tabular}
\end{table*}

\section{A Sharing Semantics Scheme Extended to Rings}\label{sec:primitive}

This section presents our sharing semantics scheme that extends the computation domain from prime fields to rings.

\subsection{Motivation and Challenges for Computation over $\mathbb{Z}_{2^l}$}

In private inference, data is represented in fixed-point format, where the last $d$ bits encode the fractional part. Fixed-point multiplication produces $2d$ fractional bits, requiring truncation (right-shifting by $d$ bits) after each multiplication to maintain precision.

Computing over the ring $\mathbb{Z}_{2^l}$ offers two advantages over prime fields $\mathbb{F}_p$. First, modulo-$2^l$ operations map directly to native CPU word operations, avoiding costly modular arithmetic over large primes. Second, truncation reduces to efficient bit-shift operations rather than division.

However, directly replacing $\mathbb{F}_p$ with $\mathbb{Z}_{2^l}$ undermines the security of MAC-based authentication. Consider the MAC relation $t_x = \alpha \cdot x$ with $x, \alpha, t_x \in \mathbb{Z}_{2^l}$. An adversary can introduce additive errors $x', \alpha', t_x'$ such that $(x + x')(\alpha + \alpha') = t_x + t_x'$. For instance, choosing $x' = 2^{l-1}$, $\alpha' = 0$, $t_x' = 0$ corrupts $x$ undetectably whenever $\alpha$ is even---an event occurring with probability $1/2$.

\subsection{Extending Computation from $\mathbb{Z}_{2^l}$ to $\mathbb{Z}_{2^{l+s}}$}

To address this vulnerability, we design a ring-based authenticated sharing scheme that separates the computation domain from the correctness domain, drawing on the domain-extension idea of SPDZ$_{2^k}$~\cite{cramer2018spd} but tailoring it to the HA-MSDM setting. Specifically, we lift all share arithmetic to $\mathbb{Z}_{2^{l+s}}$, while security and correctness remain guaranteed modulo $2^l$. The additional $s$ bits ensure that an adversary can tamper with intermediate values undetectably with probability at most $2^{-s+\lceil\log(s+1)\rceil}$.

TABLE~\ref{tab:comp_domains} specifies the computation domains for each element type and its additive shares under our scheme. Values and masks reside in $\mathbb{Z}_{2^l}$, their additive shares and all MAC-related quantities reside in $\mathbb{Z}_{2^{l+s}}$, and the MAC key $\alpha$ is shared over $\mathbb{Z}_{2^s}$. The batch check protocol $\pi_{\text{batch\_check}}$ that verifies computational integrity using this scheme is provided in Appendix~I of the supplementary material, and its security proof is provided in Appendix~II .

\begin{table}[!t]
\caption{Computation Domains for Elements and Shares\label{tab:comp_domains}}
\centering
\setlength{\tabcolsep}{3pt}
\begin{tabular}{c|c|c|c}
\hline
\textbf{Elements} & \textbf{Domain} & \textbf{Shares} & \textbf{Domain} \\
\hline
$x, m_x, \lambda_x$ & $\mathbb{Z}_{2^l}$ & $[x], [m_x], [\lambda_x]$ & $\mathbb{Z}_{2^{l+s}}$ \\
\hline
$t_x, t_{m_x}, t_{\lambda_x}$ & $\mathbb{Z}_{2^{l+s}}$ & $[t_x], [t_{m_x}], [t_{\lambda_x}]$ & $\mathbb{Z}_{2^{l+s}}$ \\
\hline
$\alpha$ & $\mathbb{Z}_{2^{l+s}}$ & $[\alpha]$ & $\mathbb{Z}_{2^s}$ \\
\hline
\end{tabular}
\end{table}

\section{Efficient MPC Layer}\label{sec:protocols}

This section introduces our efficient MPC protocols and analyze our MPC protocols' advantages over state-of-the-art MPC framework regarding time complexity, communication overhead, and communication rounds. Each protocol consists of two phases: preprocessing phase and online phase. In this section, when no ambiguity arises, we use ($=$) to denote ($\equiv_{l+s}$).

$\mathcal{F}_{\text{sh}}(\text{Rand}, P_d)$, $\mathcal{F}_{\text{sh}}(\text{Rand}, \text{HP})$ and $\mathcal{F}_{\text{sh}}(v, \text{HP})$ used in this section are defined in~\cite{karmakar2024asterisk} and also described in Section~\ref{rel:variants}. Note that, in the protocols presented below, $\pi_{\text{batch\_check}}$ may be invoked as subprotocols.

\subsection{Multiplication-with-truncation Protocol $\pi_{\text{multTrun}}$}

Private inference primarily operates on decimal values. In practice, inputs are converted from floating-point to fixed-point format (with $d$ fractional bits) at the entry of the protocol, fixed-point arithmetic is used throughout the computation, and the result is converted back to floating-point at output. A key challenge arises from multiplication: when two fixed-point values with $d$ fractional bits are multiplied, the product contains $2d$ fractional bits. Without correction, repeated multiplications cause the fractional part to progressively overflow into the integer part, corrupting the result. Therefore, the product must be truncated by right-shifting $d$ bits after each multiplication.
 
SecureML~\cite{mohassel2017secureml} first introduced the truncation pair technique, where a pair $(\lambda_z, \lambda'_z)$ satisfying $\lambda'_z = \lfloor \lambda_z / 2^d \rfloor$ is generated during preprocessing, enabling truncation via the local operation $m_{z'} = \lfloor m_z / 2^d \rfloor$ in the online phase, since the masked value and the mask are truncated simultaneously to maintain alignment. Subsequent works~\cite{yuan2024md, zhang2025md, koti2021swift} further integrate the multiplication and truncation operations into a single protocol, thereby merging their communication rounds; we refer readers to the recent SoK~\cite{harth2025sok} for a systematic treatment of truncation designs in private inference.
 
However, generating truncation pairs efficiently remains a challenge in existing frameworks. MD-ML~\cite{yuan2024md} and MD-SONIC~\cite{zhang2025md} rely on the $\mathcal{F}_{\text{edaBits}}$ functionality, while Swift~\cite{koti2021swift} constructs a dedicated truncation pair generation functionality $\mathcal{F}_{\text{TrGen}}$ tailored to its specific model. Both approaches incur substantial preprocessing overhead. Our insight is that the HA-MSDM model offers a structural advantage over prior threat models: since HP knows the mask plaintext $\lambda_z$, it can locally compute $\lambda'_z = \lfloor \lambda_z / 2^d \rfloor$ and distribute it via a single non-interactive invocation of $\mathcal{F}_{\text{sh}}(\lambda'_z, \text{HP})$. This reduces truncation pair generation from the costly $\mathcal{F}_{\text{edaBits}}$-based construction in MSDM frameworks~\cite{yuan2024md,zhang2025md} to a single element of preprocessing communication. Building on this observation, we instantiate the multiplication-with-truncation over the ring $\mathbb{Z}_{2^{l+s}}$ by combining our ring-domain authenticated sharing (Section~\ref{sec:primitive}) with HP-assisted truncation-pair generation, yielding the unified protocol $\pi_{\text{multTrun}}$ described in Fig.~\ref{proto:multTrun}.

\textbf{Proposition 1 (Correctness of $\pi_{\text{multTrun}}$).}
\textit{In the protocol $\pi_{\text{multTrun}}$, let $(xy)_s$ denote the signed
interpretation of $xy \bmod 2^l$, i.e., $(xy)_s = xy \bmod 2^l$ if
$xy \bmod 2^l < 2^{l-1}$, and $(xy)_s = (xy \bmod 2^l) - 2^l$ otherwise.
If $|(xy)_s| \leq 2^{\ell}$ for some $\ell < l - 1$, then with probability
at least $1 - 2^{\ell - l}$, the output $z'$ satisfies}
\[
z' \equiv_l \lfloor (xy)_s / 2^d \rfloor + v, \quad \text{for some } v \in \{0, 1\}.
\]

\textit{Proof.}
In $\pi_{\text{multTrun}}$, the online phase yields
$m_z \equiv_l xy + \lambda_z$, where $\lambda_z$ is uniformly random in
$\mathbb{Z}_{2^l}$. Define $w = xy \bmod 2^l$. Since
$w, \lambda_z \in [0, 2^l)$, we can write
\[
m_z = w + \lambda_z - 2^l \cdot u,
\]
where $u = 1$ if $w + \lambda_z \geq 2^l$, and $u = 0$ otherwise.
Similarly, define $v = 1$ if
$(w \bmod 2^d) + (\lambda_z \bmod 2^d) \geq 2^d$, and $v = 0$ otherwise.

Using $\lfloor a / q \rfloor = (a - (a \bmod q)) / q$ for positive
integers, we derive:
\begin{align}
z' &= \lfloor m_z / 2^d \rfloor - \lfloor \lambda_z / 2^d \rfloor
\nonumber \\
&= \lfloor w / 2^d \rfloor + \lfloor \lambda_z / 2^d \rfloor
   - 2^{l-d} u + v
   - \lfloor \lambda_z / 2^d \rfloor \nonumber \\
&= \lfloor w / 2^d \rfloor - 2^{l-d} u + v. \label{eq:trunc_core}
\end{align}

We now analyze two cases based on the sign of $(xy)_s$.

\textbf{Case 1: $(xy)_s \geq 0$.} Here $w = (xy)_s \leq 2^{\ell}$,
and the correct truncation is $\lfloor w / 2^d \rfloor$.
By~(\ref{eq:trunc_core}), this requires $u = 0$. The error event
$u = 1$ occurs when $\lambda_z \geq 2^l - w$, so
\[
\Pr[u = 1] = \frac{w}{2^l} \leq 2^{\ell - l}.
\]

\textbf{Case 2: $(xy)_s < 0$.} Here
$w = 2^l - |(xy)_s|$ and $|(xy)_s| \leq 2^{\ell}$.
By~(\ref{eq:trunc_core}), the correct result requires $u = 1$.
Write $|(xy)_s| = q \cdot 2^d + r$ with $0 \leq r < 2^d$:
\begin{itemize}
\item If $r = 0$: $\lfloor w / 2^d \rfloor = 2^{l-d} - q$,
so $z' \equiv_l -q + v \equiv_l \lfloor (xy)_s / 2^d \rfloor + v$.
\item If $r > 0$: $\lfloor w / 2^d \rfloor = 2^{l-d} - q - 1$,
so $z' \equiv_l -(q+1) + v \equiv_l \lfloor (xy)_s / 2^d \rfloor + v$.
\end{itemize}
The error event $u = 0$ occurs when $\lambda_z < |(xy)_s|$, so
\[
\Pr[u = 0] = \frac{|(xy)_s|}{2^l} \leq 2^{\ell - l}.
\]

In both cases, $z' \equiv_l \lfloor (xy)_s / 2^d \rfloor + v$ with
probability at least $1 - 2^{\ell - l}$.
\hfill$\square$

\begin{figure}[!t]
\savebox{\protobox}{%
\begin{minipage}{\dimexpr\linewidth-20pt}
\small
 
\textbf{Preprocessing phase:}
 
For a gate with inputs $\llbracket x \rrbracket$ and $\llbracket y \rrbracket$, HP knows $\lambda_x, \lambda_y$, parties know $\langle \lambda_x \rangle$ and $\langle \lambda_y \rangle$.
\begin{itemize}[leftmargin=10pt, itemsep=1pt, topsep=2pt, parsep=0pt, label=$\bullet$]
\item HP and parties execute $\mathcal{F}_{\text{sh}}(\text{Rand}, \text{HP})$ once. HP obtains the output wire mask $\lambda_z$, parties obtain $\langle \lambda_z \rangle$.
\item HP computes $\lambda_{xy} \equiv_l \lambda_x \cdot \lambda_y$. HP and parties execute $\mathcal{F}_{\text{sh}}(\lambda_{xy}, \text{HP})$ to generate $\langle \lambda_{xy} \rangle$.
\item HP computes $\lambda'_z \equiv_l \lfloor \lambda_z / 2^d \rfloor$. HP and parties execute $\mathcal{F}_{\text{sh}}(\lambda'_z, \text{HP})$ to generate $\langle \lambda'_z \rangle$.
\end{itemize}
 
\dashedline
 
\textbf{Online phase:}
 
\textbf{Input:} $P_i$ has $\llbracket x \rrbracket_i$, $\llbracket y \rrbracket_i$, $\langle \lambda_{xy} \rangle_i$, $\langle \lambda_z \rangle_i$, $\langle \lambda'_z \rangle_i$.
 
\textbf{Output:} For $i \in [1,n]$, $P_i$ outputs $\llbracket z' \rrbracket_i = (m_{z'}, \langle \lambda'_z \rangle_i)$, where $z' \equiv_l \lfloor (x \cdot y) / 2^d \rfloor$.
 
\textbf{Protocol:}
\begin{itemize}[leftmargin=10pt, itemsep=1pt, topsep=2pt, parsep=0pt, label=$\bullet$]
\item For $i \in [1,n]$, $P_i$ computes $[t_{m_z}]_i = m_x \cdot m_y \cdot [\alpha]_i - m_x \cdot [t_{\lambda_y}]_i - m_y \cdot [t_{\lambda_x}]_i + [t_{\lambda_{xy}}]_i + [t_{\lambda_z}]_i$.
\item $P_1$ computes $[m_z]_1 = m_x \cdot m_y - m_x \cdot [\lambda_y]_1 - m_y \cdot [\lambda_x]_1 + [\lambda_{xy}]_1 + [\lambda_z]_1$ and sends it to $P_{\text{king}}$.
\item For $i \in [1,n] \setminus \{1\}$, $P_i$ computes $[m_z]_i = -m_x \cdot [\lambda_y]_i - m_y \cdot [\lambda_x]_i + [\lambda_{xy}]_i + [\lambda_z]_i$ and sends it to $P_{\text{king}}$.
\item $P_{\text{king}}$ reconstructs $m_z = \sum_{i=1}^{n} [m_z]_i$ and sends $m_z$ to all parties (excluding HP).
\item For $i \in [1,n]$, $P_i$ computes $m_{z'} \equiv_l \lfloor m_z / 2^d \rfloor$ and outputs $\llbracket z' \rrbracket_i = (m_{z'}, \langle \lambda'_z \rangle_i)$.
\end{itemize}
 
\end{minipage}%
}%
\centering
\begin{tikzpicture}
\node[inner sep=8pt, outer sep=0pt] (box) {\usebox{\protobox}};
\draw[line width=1pt] (box.north west) rectangle (box.south east);
\node[draw, line width=0.8pt, fill=white, inner sep=3pt, rounded corners=3pt] at ([xshift=42pt]box.north west) {\small\textbf{Protocol} $\pi_{\text{multTrun}}$};
\end{tikzpicture}
\caption{Multiplication-with-truncation protocol.}
\label{proto:multTrun}
\end{figure}

\subsection{Integer-polynomial Protocol $\pi_{\text{poly\_integer}}$}

 

A fundamental building block for polynomial evaluation in MPC is computing the power series $\llbracket x \rrbracket, \llbracket x^2 \rrbracket, \ldots, \llbracket x^k \rrbracket$ from a shared input $\llbracket x \rrbracket$. The conventional approach organizes $k-1$ sequential multiplications into a balanced binary tree~\cite{knott2021crypten, patra2021aby2}, reducing round complexity from $O(k)$ to $O(\log k)$. However, each multiplication still requires an interactive reconstruction, so the total online communication scales as $O(nk)$ ring elements per evaluation.

We propose an integer polynomial evaluation protocol $\pi_{\text{poly\_integer}}$ whose online phase is dominated by a single $P_{\text{king}}$ reconstruction, with $2$ fixed communication rounds and $2(n-1)$ ring elements total, \emph{independent of both $k$ and the number of parties $n$}. This yields two concrete advantages over the binary tree baseline that are not captured by the asymptotic round count alone:
\begin{itemize}[leftmargin=12pt,itemsep=1pt,topsep=2pt,parsep=0pt]
\item \textbf{Communication}, not rounds, is the dominant savings. For typical CNN activations with $k=4$, the binary tree needs $\log_2 k = 2$ rounds---matching our round count---but still incurs $O(nk)$ online communication per evaluation, versus our $O(n)$. On millions of activations in a single inference, this difference is what drives the measured WAN speedup (Section~\ref{sec:experiments}).
\item \textbf{Scalability in $n$.} The binary tree's per-multiplication reconstruction is $O(n)$ elements among parties, so its total online communication is $O(n^2 k)$ across all parties, whereas ours remains $O(n)$. This gap widens as $n$ grows.
\end{itemize}
The savings come from Parts~I--III of the online phase (Fig.~\ref{proto:poly_integer}) being entirely local: only Part~IV requires two rounds for $P_{\text{king}}$ reconstruction.
 
The core idea is to transform the computation of $\llbracket x^i \rrbracket$ into computing $\llbracket x^{i-j} \cdot r^j \rrbracket$ through factorization, where $r$ is a random value generated by HP during preprocessing. The key observation is the recurrence
\begin{equation}\label{eq:exp_iter}
x^{i-j} \cdot r^j = (x - r) \cdot x^{i-j-1} \cdot r^j + x^{i-j-1} \cdot r^{j+1},
\end{equation}
where $c = x - r$ is a public constant known to all parties (revealed in Part~I) but kept secret from HP. Since $c$ is a public scalar, the operation $c \cdot \llbracket x^{i-j-1} \cdot r^j \rrbracket + \llbracket x^{i-j-1} \cdot r^{j+1} \rrbracket$ requires no interaction---it is a local linear combination on shares. The base cases $\llbracket r^i \rrbracket$ are prepared during preprocessing. By applying Equation~(\ref{eq:exp_iter}) via dynamic programming, all parties simultaneously obtain $\llbracket x^j \rrbracket = \llbracket x^j \cdot r^0 \rrbracket$ for $j \in [1,k]$ without any online communication. The polynomial output $\llbracket y \rrbracket$ is then computed as a local linear combination of these terms and reconstructed via $P_{\text{king}}$ in two rounds. The protocol $\pi_{\text{poly\_integer}}$ is described in Fig.~\ref{proto:poly_integer}.
 
\textbf{Proposition 2 (Correctness of $\pi_{\text{poly\_integer}}$).}
\textit{In the protocol $\pi_{\text{poly\_integer}}$, the output satisfies $m_y \equiv_l y + \lambda_y$, where $y = \sum_{j=0}^{k} a_j \cdot x^j$, and $\sum_{i=1}^{n}[t_{m_y}]_i = \alpha \cdot m_y$.}
 
\textit{Proof.}
We establish correctness in three steps, corresponding to Parts~I--III of the online phase.
 
\textit{Step 1 (Part I).} Each party computes $c = m_x - \delta = (x + \lambda_x) - (\lambda_x + r) = x - r$. Since $m_x$ is public among parties (from the preceding gate's $P_{\text{king}}$ broadcast) and $\delta$ is broadcast by the semi-honest HP, all parties obtain the correct value $c$.
 
\textit{Step 2 (Part II).} We prove by induction that $\text{arr}[i][j] = \llbracket x^{i-j} \cdot r^j \rrbracket$. The base cases hold by construction: $\text{arr}[0][i] = \llbracket r^i \rrbracket = \llbracket x^0 \cdot r^i \rrbracket$ and $\text{arr}[1][0] = \llbracket x \rrbracket = \llbracket x^1 \cdot r^0 \rrbracket$. For the inductive step, assuming $\text{arr}[i-1][j]$ and $\text{arr}[i-1][j+1]$ are correct:
\begin{align*}
\text{arr}[i][j] &= c \cdot \llbracket x^{i\!-\!1\!-\!j} \!\cdot\! r^j \rrbracket + \llbracket x^{i\!-\!1\!-\!j} \!\cdot\! r^{j+1} \rrbracket \\
&= \llbracket (x\!-\!r) \cdot x^{i\!-\!1\!-\!j} \!\cdot\! r^j + x^{i\!-\!1\!-\!j} \!\cdot\! r^{j+1} \rrbracket = \llbracket x^{i-j} \!\cdot\! r^j \rrbracket.
\end{align*}
Here, $c$ is a public scalar, so the linear combination operates identically on all three components $(m, [\lambda], [t_\lambda])$ of each $\llbracket \cdot \rrbracket$ entry, preserving the sharing structure. In particular, $\text{arr}[j][0] = \llbracket x^j \rrbracket$ for all $j \in [1,k]$.
 
\textit{Step 3 (Parts III--IV).} Let $S = a_0 + \sum_{j=1}^{k} a_j \cdot m_{x^j}$. The shares satisfy:
\begin{align*}
\sum_{i=1}^{n} [m_y]_i &= S + \sum_{i=1}^{n}[\lambda_y]_i - \sum_{j=1}^{k} a_j \sum_{i=1}^{n}[\lambda_{x^j}]_i \\
&= a_0 + \sum_{j=1}^{k} a_j (x^j\! +\! \lambda_{x^j}) + \lambda_y - \sum_{j=1}^{k} a_j \lambda_{x^j} \\
&= \sum_{j=0}^{k} a_j \cdot x^j + \lambda_y = y + \lambda_y = m_y.
\end{align*}
The MAC shares satisfy:
\begin{align*}
\sum_{i=1}^{n} [t_{m_y}]_i &= S \cdot \alpha - \sum_{j=1}^{k} a_j \cdot \alpha \lambda_{x^j} + \alpha \lambda_y = \alpha \cdot m_y.
\end{align*}
Hence $P_{\text{king}}$ correctly reconstructs $m_y = y + \lambda_y$, and the MAC relation holds for $\pi_{\text{batch\_check}}$. \hfill$\square$

\begin{figure}[!t]
\savebox{\protobox}{%
\begin{minipage}{\dimexpr\linewidth-20pt}
\small
 
\textbf{Preprocessing phase:}
 
For a gate with input $\llbracket x \rrbracket$ and public polynomial coefficients $(a_0, a_1, \ldots, a_k)$, output $\llbracket y \rrbracket$, where $y \equiv_l \sum_{i=0}^{k} a_i \cdot x^i$. HP knows $\lambda_x$, parties know $\langle \lambda_x \rangle$.
\begin{itemize}[leftmargin=10pt, itemsep=1pt, topsep=2pt, parsep=0pt, label=$\bullet$]
\item HP and parties execute $k$ times of $\mathcal{F}_{\text{sh}}(\text{Rand}, \text{HP})$. HP obtains $\lambda_{r}, \lambda_{r^2}, \ldots, \lambda_{r^k}$ and parties obtain $\langle \lambda_{r} \rangle, \langle \lambda_{r^2} \rangle, \ldots, \langle \lambda_{r^k} \rangle$.
\item HP instantiates a random number $r$ and computes $r^2, r^3, \ldots, r^k \xleftarrow{\$} \mathbb{Z}_{2^{l}}$  and the masked values $m_{r^i} \equiv_l r^i + \lambda_{r^i}$ for $i \in [1,k]$, and sends $m_{r^i}$ to all parties.
\item Parties reconstruct $\llbracket r^i \rrbracket = (m_{r^i}, \langle \lambda_{r^i} \rangle)$ for $i \in [1,k]$.
\item HP and parties execute $\mathcal{F}_{\text{sh}}(\text{Rand}, \text{HP})$ once. HP obtains the output wire mask $\lambda_y$, parties obtain $\langle \lambda_y \rangle$.
\item HP computes $\delta \equiv_l \lambda_x + r$ and sends $\delta$ to all parties.
\end{itemize}
 
\dashedline
 
\textbf{Online phase:}
 
\textbf{Input:} For $i \in [1,n]$, $P_i$ has $\llbracket x \rrbracket_i$, $\llbracket r^j \rrbracket_i$ for $j \in [1,k]$, $\langle \lambda_y \rangle_i$, and $\delta$. HP has $\lambda_x, r, \lambda_{r^j}, \lambda_y$.
 
\textbf{Output:} For $i \in [1,n]$, $P_i$ outputs $\llbracket y \rrbracket_i = (m_y, \langle \lambda_y \rangle_i)$ and $[t_{m_y}]_i$, where $m_y \equiv_l y + \lambda_y$, $y = \sum_{j=0}^{k} a_j \cdot x^j$.
 
\textit{Part I: Reveal $c = x - r$ \textnormal{(local, 0 rounds)}}
\begin{itemize}[leftmargin=10pt, itemsep=1pt, topsep=2pt, parsep=0pt, label=$\bullet$]
\item For $i \in [1,n]$, $P_i$ locally computes $c \equiv_l m_x - \delta \equiv_l x - r$.
\end{itemize}
 
\textit{Part II: Compute $\llbracket x^j \rrbracket$ via DP \textnormal{(local, 0 rounds)}}
\begin{itemize}[leftmargin=10pt, itemsep=1pt, topsep=2pt, parsep=0pt, label=$\bullet$]
\item All parties initialize $\text{arr}[k\!+\!1][k\!+\!1]$, where $\text{arr}[i][j]$ stores $\llbracket x^{i-j} \cdot r^j \rrbracket$. Each entry contains three components $(m, [\lambda], [t_\lambda])$ processed via identical linear operations. Execute:
 
\hspace{6pt} $\text{for } (i\!=\!1;\ i \!\leq\! k;\ i\text{++}) \{ \text{arr}[0][i] \!=\! \llbracket r^i \rrbracket \}$;\quad $\text{arr}[1][0] \!=\! \llbracket x \rrbracket$;
 
\hspace{6pt} $\text{for } (i\!=\!1;\ i \!<\! k;\ i\text{++})$\quad $\text{for } (j\!=\!1;\ j \!\leq\! k\!-\!i;\ j\text{++})$
 
\hspace{18pt} $\text{arr}[i][j] = c \cdot \text{arr}[i\!-\!1][j] + \text{arr}[i\!-\!1][j\!+\!1]$;
 
\hspace{6pt} $\text{for } (i\!=\!2;\ i \!\leq\! k;\ i\text{++})$\quad $\text{arr}[i][0] = c \cdot \text{arr}[i\!-\!1][0] + \text{arr}[i\!-\!1][1]$;
 
\item All parties extract $\llbracket x^j \rrbracket = \text{arr}[j][0]$ for $j \in [1,k]$, obtaining $m_{x^j}$, $[\lambda_{x^j}]_i$, $[t_{\lambda_{x^j}}]_i$.
\end{itemize}
 
\textit{Part III: Polynomial evaluation \textnormal{(local, 0 rounds)}}
\begin{itemize}[leftmargin=10pt, itemsep=1pt, topsep=2pt, parsep=0pt, label=$\bullet$]
\item Let $S \equiv_l a_0 + \sum_{j=1}^{k} a_j \cdot m_{x^j}$. For $i \in [1,n]$, $P_i$ locally computes:
 
\hspace{6pt} $[m_y]_i = [\lambda_y]_i - \sum_{j=1}^{k} a_j \cdot [\lambda_{x^j}]_i$;\quad if $i = 1$: $[m_y]_1 \mathrel{+}= S$.
 
\hspace{6pt} $[t_{m_y}]_i = S \cdot [\alpha]_i - \sum_{j=1}^{k} a_j \cdot [t_{\lambda_{x^j}}]_i + [t_{\lambda_y}]_i$.
\end{itemize}
 
\textit{Part IV: $P_{\textnormal{king}}$ reconstruction \textnormal{(2 rounds)}}
\begin{itemize}[leftmargin=10pt, itemsep=1pt, topsep=2pt, parsep=0pt, label=$\bullet$]
\item For $i \in [1,n] \setminus \{P_{\text{king}}\}$, $P_i$ sends $[m_y]_i$ to $P_{\text{king}}$.
\item $P_{\text{king}}$ reconstructs $m_y = \sum_{i=1}^{n} [m_y]_i$ and sends $m_y$ to all parties (excluding HP).
\item For $i \in [1,n]$, $P_i$ outputs $\llbracket y \rrbracket_i = (m_y, \langle \lambda_y \rangle_i)$ and stores $[t_{m_y}]_i$ for subsequent $\pi_{\text{batch\_check}}$.
\end{itemize}
 
\end{minipage}%
}%
\centering
\begin{tikzpicture}
\node[inner sep=8pt, outer sep=0pt] (box) {\usebox{\protobox}};
\draw[line width=1pt] (box.north west) rectangle (box.south east);
\node[draw, line width=0.8pt, fill=white, inner sep=3pt, rounded corners=3pt] at ([xshift=48pt]box.north west) {\small\textbf{Protocol} $\pi_{\text{poly\_integer}}$};
\end{tikzpicture}
\caption{Integer polynomial evaluation protocol.}
\label{proto:poly_integer}
\end{figure}

\subsection{Fixed-point Polynomial Protocol $\pi_{\text{poly\_fixed}}$}

In private inference, polynomial coefficients and inputs are real-valued and represented in fixed-point format with $d$ fractional bits. A direct application of $\pi_{\text{poly\_integer}}$ to fixed-point operands would accumulate $dk$ fractional bits after evaluating a degree-$k$ polynomial, corrupting the result. Moreover, since polynomial-based activation function approximation can only guarantee high-precision fitting within a fixed interval~\cite{garimella2021sisyphus,diaa2024fast}, our goal is to design an efficient fixed-point polynomial protocol that guarantees computational correctness when the input is bounded within a predetermined interval. The fixed-point polynomial protocol $\pi_{\text{poly\_fixed}}$ achieves this by absorbing the scaling into integer coefficients, enforcing an explicit precondition on the input range, and applying a single truncation at the end.

Concretely, given real-valued coefficients $(a'_0, a'_1, \ldots, a'_k)$ and a fixed-point input $x$ with $d$ fractional bits satisfying $|x| \leq q \cdot 2^d$, we define integer coefficients $b_i = a_i \cdot 2^{d(k-i)}$ where $a_i = \lfloor a'_i \cdot 2^d \rfloor$, and compute the integer polynomial $Y_{\text{int}} = \sum_{i=0}^{k} b_i \cdot x^i$ entirely over $\mathbb{Z}_{2^l}$. The final result is obtained by truncating $y' \equiv_l \lfloor Y_{\text{int}} / 2^{dk} \rfloor$, which recovers the correct fixed-point output.

To ensure correctness, the bit length $l$ must satisfy the precondition $\sum_{i=0}^{k} |b_i| \cdot (q \cdot 2^d)^i < 2^{l-1}$, so that $Y_{\text{int}}$ does not overflow $\mathbb{Z}_{2^l}$. The bound $q$ determines the valid input interval $[-q, q]$ in real-valued terms, and its selection together with other related hyperparameters will be discussed in detail in Section~\ref{sec:inference}. The protocol $\pi_{\text{poly\_fixed}}$ is described in Fig.~\ref{proto:poly_fixed}.

\textbf{Proposition 3 (Correctness of $\pi_{\text{poly\_fixed}}$).}
\textit{In the protocol $\pi_{\text{poly\_fixed}}$, let $Y_{\text{int}} = \sum_{i=0}^{k} b_i \cdot x^i$. If the precondition $|Y_{\text{int}}| < 2^{l-1}$ holds, then the output satisfies $y' \equiv_l \lfloor Y_{\text{int}} / 2^{dk} \rfloor + \varepsilon$ for some $\varepsilon \in \{0, 1\}$, with probability at least $1 - 2^{\ell' - l}$ where $\ell' = \lceil \log_2 |Y_{\text{int}}| \rceil$.}

\textit{Proof.}
Parts~I--III follow identically from Proposition~2 with $a_j$ replaced by $b_j$. After $P_{\text{king}}$ reconstruction, $m_Y \equiv_l Y_{\text{int}} + \lambda_Y$. The truncation step $m_{y'} = \lfloor m_Y / 2^{dk} \rfloor$ combined with the preprocessed truncation pair $\lambda'_Y = \lfloor \lambda_Y / 2^{dk} \rfloor$ yields $y' = m_{y'} - \lambda'_Y \equiv_l \lfloor Y_{\text{int}} / 2^{dk} \rfloor + \varepsilon$ by the same carry analysis as in Proposition~1. \hfill$\square$

\begin{figure}[!t]
\savebox{\protobox}{%
\begin{minipage}{\dimexpr\linewidth-20pt}
\small

\textbf{Precondition:} $l$ is chosen such that $\sum_{i=0}^{k} |b_i| \cdot (q \cdot 2^d)^i < 2^{l-1}$, where $b_i = a_i \cdot 2^{d(k-i)}$, $a_i = \lfloor a'_i \cdot 2^d \rfloor$, ensuring $|Y_{\text{int}}| < 2^{l-1}$.

\vspace{4pt}

\textbf{Preprocessing phase:}

For a gate with input $\llbracket x \rrbracket$ (where $|x| \leq q \cdot 2^d$) and public real-valued polynomial coefficients $(a'_0, a'_1, \ldots, a'_k)$, output $\llbracket y' \rrbracket$ where $y' \equiv_l \lfloor Y_{\text{int}} / 2^{dk} \rfloor$, $Y_{\text{int}} = \sum_{i=0}^{k} b_i \cdot x^i$.
\begin{itemize}[leftmargin=10pt, itemsep=1pt, topsep=2pt, parsep=0pt, label=$\bullet$]
\item Execute the same preprocessing steps as $\pi_{\text{poly\_integer}}$ (Fig.~\ref{proto:poly_integer}): generate $\llbracket r^j \rrbracket$ for $j \in [1,k]$, output mask $\lambda_Y$, and $\delta \equiv_l \lambda_x + r$.
\item \textbf{(Additional)} HP computes the truncation pair $\lambda'_Y = \lfloor \lambda_Y / 2^{dk} \rfloor$. HP and parties execute $\mathcal{F}_{\text{sh}}(\lambda'_Y, \text{HP})$ to generate $\langle \lambda'_Y \rangle$.
\end{itemize}

\dashedline

\textbf{Online phase:}

\textbf{Input:} Same as $\pi_{\text{poly\_integer}}$, plus $\langle \lambda'_Y \rangle_i$ for each $P_i$.

\textbf{Output:} For $i \in [1,n]$, $P_i$ outputs $\llbracket y' \rrbracket_i = (m_{y'}, \langle \lambda'_Y \rangle_i)$ and $[t_{m_Y}]_i$, where $y' \equiv_l \lfloor Y_{\text{int}} / 2^{dk} \rfloor$.

\textit{Parts I--II: Identical to $\pi_{\text{poly\_integer}}$ \textnormal{(local, 0 rounds)}}
\begin{itemize}[leftmargin=10pt, itemsep=1pt, topsep=2pt, parsep=0pt, label=$\bullet$]
\item Compute $c \equiv_l x - r$ and derive $\llbracket x^j \rrbracket$ for $j \in [1,k]$ via the same DP procedure as in Fig.~\ref{proto:poly_integer}.
\end{itemize}

\textit{Part III: Integer polynomial evaluation \textnormal{(local, 0 rounds)}}
\begin{itemize}[leftmargin=10pt, itemsep=1pt, topsep=2pt, parsep=0pt, label=$\bullet$]
\item Identical to Part~III of $\pi_{\text{poly\_integer}}$, with each $a_j$ replaced by $b_j = a_i \cdot 2^{d(k-i)}$.
\end{itemize}

\textit{Part IV: $P_{\textnormal{king}}$ reconstruction and truncation \textnormal{(2 rounds)}}
\begin{itemize}[leftmargin=10pt, itemsep=1pt, topsep=2pt, parsep=0pt, label=$\bullet$]
\item For $i \in [1,n] \setminus \{P_{\text{king}}\}$, $P_i$ sends $[m_Y]_i$ to $P_{\text{king}}$.
\item $P_{\text{king}}$ reconstructs $m_Y \equiv_l \sum_{i=1}^{n} [m_Y]_i$ and sends $m_Y$ to all parties (excluding HP).
\item \textbf{(Truncation)} For $i \in [1,n]$, $P_i$ computes $m_{y'} \equiv_l \lfloor m_Y / 2^{dk} \rfloor$ and outputs $\llbracket y' \rrbracket_i = (m_{y'}, \langle \lambda'_Y \rangle_i)$.
\item For $i \in [1,n]$, $P_i$ stores $[t_{m_Y}]_i$ for subsequent $\pi_{\text{batch\_check}}$.
\end{itemize}

\end{minipage}%
}%
\centering
\begin{tikzpicture}
\node[inner sep=8pt, outer sep=0pt] (box) {\usebox{\protobox}};
\draw[line width=1pt] (box.north west) rectangle (box.south east);
\node[draw, line width=0.8pt, fill=white, inner sep=3pt, rounded corners=3pt] at ([xshift=42pt]box.north west) {\small\textbf{Protocol} $\pi_{\text{poly\_fixed}}$};
\end{tikzpicture}
\caption{Fixed-point polynomial evaluation protocol.}
\label{proto:poly_fixed}
\end{figure}

\section{High-Accuracy CNN Inference Layer}\label{sec:inference}

The inference layer implements all operations required for CNN private inference, including convolutional layers, fully connected layers, average pooling layers, max pooling layers, batch normalization (BN) layers, and activation function layers. Among these, convolutional and fully connected layers are linear operations implemented via straightforward extensions of $\pi_{\text{multTrun}}$ to matrix and convolution forms. Average pooling is realized through fixed-point scalar multiplication-truncation, and max pooling is computed via $\text{Max}(a,b) = \text{ReLU}(a-b) + b$. BN layers serve a dual role: accelerating training convergence and constraining activation inputs to the valid interval required by $\pi_{\text{poly\_fixed}}$. The implementation details of these layers are provided in Appendix~III of the supplementary material.

This section focuses on two core problems that determine the accuracy of private inference: (1) how to achieve high-precision polynomial activation without sacrificing model expressiveness (Section~\ref{subsec:activation_strategy}), and (2) how to select hyperparameters such that the correctness conditions of $\pi_{\text{poly\_fixed}}$ are satisfied with negligible failure probability (Section~\ref{subsec:hyperparameter}).

\subsection{High-Accuracy Polynomial Activation Strategy}\label{subsec:activation_strategy}

A common approach to computing nonlinear activation functions in MPC-based private inference is polynomial approximation via least squares fitting~\cite{gilad2016cryptonets,garimella2021sisyphus,diaa2024fast}. This method constructs a polynomial that minimizes the mean squared error (MSE) against the target activation function over a prescribed interval. However, outside this interval the polynomial diverges rapidly from the original function, a phenomenon known as the \textit{escaping activation problem}~\cite{garimella2021sisyphus,diaa2024fast}.

To analyze the end-to-end accuracy impact, we decompose the total inference accuracy loss into two orthogonal components:
\begin{itemize}
\item \textbf{Gap 1 (Approximation Gap):} $\text{Acc}(\text{ReLU}, \text{plain}) - \text{Acc}(\text{Poly}, \text{plain})$, caused by replacing ReLU with polynomial approximation, reflecting the loss of model expressiveness.
\item \textbf{Gap 2 (Computation Gap):} $\text{Acc}(\text{Poly}, \text{plain}) - \text{Acc}(\text{Poly}, \text{cipher})$, caused by fixed-point arithmetic and truncation errors during secure computation.
\end{itemize}

The state-of-the-art work PILLAR~\cite{diaa2024fast} effectively addresses Gap~2 through activation regularization, which encourages the model to keep activation inputs within the fitting interval during training. However, as shown in TABLE~\ref{tab:accuracy_gap}, PILLAR's regularization simultaneously constrains the model's intrinsic expressiveness, leaving a substantial Gap~1 of approximately 2--5\%. Our goal is to close Gap~1 while preserving PILLAR's effectiveness on Gap~2, thereby minimizing the total loss $\text{Acc}(\text{ReLU}, \text{plain}) - \text{Acc}(\text{Poly}, \text{cipher})$.

Our strategy builds upon PILLAR's foundation (Sections~\ref{sssec:qapf}--\ref{sssec:reg_clip}) and introduces two complementary techniques---knowledge distillation and warm starting (Sections~\ref{sssec:kd}--\ref{sssec:warm})---to recover the expressiveness lost due to regularization.

\paragraph{\textbf{Quantization-Aware Polynomial Fitting}}\label{sssec:qapf}

Standard least squares fitting produces real-valued coefficients that incur additional quantization error when converted to fixed-point representation. Following PILLAR~\cite{diaa2024fast}, we perform the fitting directly over the discretized fixed-point domain, ensuring that the resulting coefficients $a_i = \lfloor a'_i \cdot 2^d \rfloor$ are exactly representable. These coefficients are then transformed into the integer coefficients $b_i = a_i \cdot 2^{d(k-i)}$ required by $\pi_{\text{poly\_fixed}}$. The specific fitting interval, polynomial degree, and coefficient values are detailed in Section~\ref{subsec:hyperparameter}.

\paragraph{\textbf{Activation Regularization and Clipping}}\label{sssec:reg_clip}

Activation Regularization. To mitigate the escaping activation problem during training, we adopt the activation regularization framework of PILLAR~\cite{diaa2024fast}. A penalty term is added to the training loss that penalizes activation inputs falling outside the polynomial fitting interval, encouraging the model to learn representations whose intermediate activations remain within the valid range. We additionally explored three alternative regularization formulations; however, their performance did not surpass the PILLAR formulation under our experimental settings. These alternatives are described in Appendix~IV of the supplementary material.

Clipping. In early training stages, activation inputs may significantly exceed the fitting interval before the regularization has taken effect, leading to extreme polynomial outputs and numerical divergence. To stabilize training, we apply element-wise clipping to activation outputs after the regularization penalty is computed, ensuring that clipping does not interfere with the penalty gradients. Clipping is used exclusively during training and is removed at inference time.

\paragraph{\textbf{Knowledge Distillation from ReLU Teacher}}\label{sssec:kd}

Activation regularization effectively constrains activation inputs to the fitting interval, but this comes at a cost: the model must satisfy the regularization constraint in addition to optimizing classification accuracy, which limits its representational capacity. As shown in TABLE~\ref{tab:accuracy_gap}, this leads to a Gap~1 of approximately 2--5\%. To recover this lost accuracy, we introduce knowledge distillation (KD) from a pre-trained ReLU teacher model.

The ReLU teacher is trained without any polynomial constraint and thus achieves the full accuracy of the original architecture. We use its output to guide the polynomial student during training. The student's training loss is:
\[\label{eq:kd_loss}
\mathcal{L} = (1 - \alpha) \cdot \mathcal{L}_{\text{CE}}(\mathbf{y}_s, \mathbf{y}_{\text{true}}) + \alpha \cdot T^2 \cdot D_{\text{KL}}\!\left(\sigma\!\left(\frac{\mathbf{z}_s}{T}\right) \middle\| \sigma\!\left(\frac{\mathbf{z}_t}{T}\right)\right),
\]
where $\mathbf{z}_s$ and $\mathbf{z}_t$ denote the logits of the student and teacher respectively, $\sigma(\cdot)$ is the softmax function, and the factor $T^2$ compensates for the gradient magnitude reduction caused by temperature scaling~\cite{hinton2015distilling}. The hyperparameter $\alpha \in [0,1]$ balances the two loss terms: the cross-entropy loss $\mathcal{L}_{\text{CE}}$ against ground-truth labels, and the KL divergence against the teacher's softened output distribution. The temperature $T > 1$ controls the smoothness of the teacher's output---a higher $T$ produces softer probability distributions that reveal more information about inter-class relationships. The specific values of $\alpha$ and $T$ are discussed in Section~\ref{subsec:hyperparameter}.

The reason KD is effective for our setting is straightforward. Without KD, the polynomial student can only learn from one-hot ground-truth labels, which provide no information about relationships between classes, such as a ``cat'' image is more similar to ``dog'' than to ``truck''. The teacher's soft labels encode precisely this information. By learning from these richer targets, the polynomial student can achieve higher accuracy even under the regularization constraint, because the soft labels reduce the number of training samples needed to learn the same decision boundaries.

\paragraph{\textbf{Warm Starting from ReLU Weights}}\label{sssec:warm}

We initialize the polynomial student from the pre-trained ReLU model's weights rather than from random initialization. All convolutional and fully connected layer weights are directly copied, along with BN parameters. 

The primary motivation is training stability. With random initialization, the activation inputs at the beginning of training are widely and unpredictably distributed, causing most inputs to fall far outside the polynomial fitting interval. This triggers large regularization penalties whose gradients dominate and conflict with the classification loss gradients, making optimization difficult. In contrast, a ReLU model's learned weights already produce activation inputs that are naturally concentrated near zero. When these weights are used to initialize the polynomial student, the activation input distribution is well-behaved from the first training step, and the regularization penalty starts small. This allows the classification loss and the KD loss to guide the early training, with regularization acting as a gentle corrective force rather than a dominant one.

A secondary benefit is improved synergy with knowledge distillation. Since the student starts with weights close to the teacher's, its initial output distribution already resembles the teacher's soft labels. This makes the KD loss more informative from the outset, leading to faster and more stable convergence.

\subsection{Hyperparameter Selection and Correctness Analysis}\label{subsec:hyperparameter}

\textbf{Hyperparameter selection.} TABLE~\ref{tab:hyperparams} summarizes the hyperparameters used throughout our framework. We employ a degree-4 polynomial to approximate ReLU over the interval $[-7, 7]$ ($q = 7$), with $d = 12$ fractional bits. The polynomial coefficients, obtained via quantization-aware least squares fitting, are:
$\label{eq:poly_relu}
P(x) = -0.001220703125 x^4 + 0.1181640625 x^2 + 0.5 x + 0.40625,
$
achieving an MSE of $\text{0.016}$ against ReLU on $[-7, 7]$. For $\pi_{\text{poly\_fixed}}$, we set the ring bit length $l = 88$ and the statistical security parameter $s = 40$, so that all share arithmetic is performed over $\mathbb{Z}_{2^{128}}$. For activation regularization, the penalty threshold is $\lambda_{\text{reg}} = 6.3$ (slightly below $q = 7$ to provide a safety margin), with the penalty exponent $\gamma$ progressively increasing from 2 to 10 and the penalty weight $\beta$ from $10^{-5}$ to $2 \times 10^{-3}$ over the 120 training epochs via a piecewise schedule. For knowledge distillation, we set $\alpha = 0.7$ and $T = 4.0$. For small-capacity networks such as MiniONN, we reduce $\alpha$ to 0.05, as we empirically found that strong distillation guidance overwhelms the limited representational capacity of these models, degrading rather than improving accuracy. Training uses SGD with a learning rate of 0.013, momentum 0.9, weight decay $5 \times 10^{-4}$, and a warmup-cosine learning rate schedule with 5 warmup epochs.

\begin{table}[!t]
\caption{Summary of Hyperparameters\label{tab:hyperparams}}
\centering
\begin{tabular}{l|l}
\hline
\textbf{Parameter} & \textbf{Value} \\
\hline
Polynomial degree $k$ & 4 \\
Fitting interval $[-q, q]$ & $[-7, 7]$ \\
Fractional bits $d$ & 12 \\
Ring bit length $l$ & 88 \\
Security parameter $s$ & 40 \\
Regularization threshold $\lambda_{\text{reg}}$ & 6.3 \\
Regularization exponent $\gamma$ & $2 \to 10$ (scheduled) \\
Regularization weight $\beta$ & $10^{-5} \to 2\!\times\!10^{-3}$ (scheduled) \\
KD balance $\alpha$ & 0.7 \\
KD temperature $T$ & 4.0 \\
\hline
\end{tabular}
\end{table}

\textbf{Correctness analysis of $\pi_{\text{poly\_fixed}}$.} We verify that the chosen parameters satisfy the precondition of Proposition~3. The integer coefficients are $b_i = a_i \cdot 2^{d(k-i)}$, where $a_i = \lfloor a'_i \cdot 2^{12} \rfloor$. At the worst-case input $x_{\text{real}} = 7$, the polynomial evaluates to $P(7) \approx 7.295$, giving $|Y_{\text{int}}| = 2^{dk} \cdot |P(7)| \approx 2^{48} \times 7.295 \approx 2^{50.9}$. Since $2^{50.9} \ll 2^{87} = 2^{l-1}$, the precondition $|Y_{\text{int}}| < 2^{l-1}$ is satisfied with substantial margin. By Proposition~3, the single-invocation truncation failure probability is at most $|Y_{\text{int}}| / 2^l \leq 2^{50.9 - 88} \approx 2^{-37}$. Applying a union bound over all activation function invocations in a complete inference pass, the worst-case network-level failure probability is at most $N \cdot 2^{-37}$, where $N$ denotes the total number of activations. 

We note that this is a pessimistic upper bound, since it assumes worst-case activation inputs at $|x| = 7$. Our empirical measurements show that regularized activation inputs approximately follow a Laplace distribution centered at zero (see Appendix~V-A of the supplementary material), with the vast majority of inputs satisfying $|x| < 3$, for which $|P(x)| < 3$ and the per-invocation failure probability drops to $\approx 2^{-39}$. To empirically validate these theoretical bounds, we ran $N_{\text{run}} = 1000$ independent end-to-end ciphertext inferences of ResNet-18 on CIFAR-10 (involving roughly $1.7 \times 10^7$ cumulative $\pi_{\text{poly\_fixed}}$ invocations across the trial). \emph{Zero truncation failures were observed.} By the Clopper--Pearson one-sided bound at 95\% confidence, this corresponds to a per-inference failure rate upper bounded by $3.0 \times 10^{-3}$, which is consistent with our network-level theoretical estimate and orders of magnitude below the pessimistic union-bound worst case.

\textbf{Accuracy evaluation.} TABLE~\ref{tab:accuracy_gap} presents the plaintext inference accuracy across five CNN architectures on CIFAR-10, CIFAR-100 and ImageNet. We compare three configurations: ReLU plaintext inference (the accuracy ceiling), polynomial plaintext inference with regularization only (corresponding to the PILLAR~\cite{diaa2024fast} baseline), and polynomial plaintext inference with our full strategy (regularization + KD + warm start). The results demonstrate that regularization alone leaves a significant Gap~1 of 2.1--5.3\%, whereas our full strategy reduces this gap to within 0.2\% across all tested configurations. In most cases, our polynomial models match the ReLU baseline exactly. Comprehensive ablation studies and ciphertext inference results are presented in Section~\ref{sec:experiments}.

\begin{table}[!t]
\caption{Plaintext Inference Accuracy (\%) with Degree-4 Polynomial Approximation\label{tab:accuracy_gap}}
\centering
\begin{tabular}{l|c|c|c|c}
\hline
\textbf{Model} & \textbf{Dataset} & \makecell{\textbf{ReLU}\\\textbf{Plain}} & \makecell{\textbf{Poly Plain}\\\textbf{(Reg. Only)}} & \makecell{\textbf{Poly Plain}\\\textbf{(Ours)}} \\
\hline
ResNet-18 & CIFAR-10 & 94.7 & 92.1 & \textbf{94.7} \\
ResNet-18 & CIFAR-100 & 75.8 & 70.5 & \textbf{75.8} \\
ResNet-32 & CIFAR-10 & 94.0 & 90.9 & \textbf{94.0} \\
ResNet-32 & CIFAR-100 & 69.2 & 65.0 & \textbf{69.2} \\
ResNet-110 & CIFAR-10 & 93.4 & 91.3 & \textbf{93.4} \\
VGG-16 & CIFAR-10 & 92.6 & 90.0 & \textbf{92.6} \\
VGG-16 & CIFAR-100 & 74.2 & 72.0 & \textbf{74.0} \\
MiniONN & CIFAR-10 & 86.0 & 84.0 & \textbf{86.0} \\
ResNet-18 & ImageNet & 69.8 & 65.6 & \textbf{69.8} \\
\hline
\end{tabular}
\end{table}

\section{Security Analysis}\label{sec:security}

We analyze the security of our protocols under the HA-MSDM threat model: a non-colluding adversary either maliciously corrupts up to $n-1$ parties, or semi-honestly corrupts HP. Both $\pi_{\text{multTrun}}$ and $\pi_{\text{poly\_fixed}}$ follow Asterisk's~\cite{karmakar2024asterisk} design paradigm, where the preprocessing phase is inherently error-free (only HP sends messages), online-phase correctness is ensured by $\pi_{\text{batch\_check}}$, and fairness is guaranteed by HP releasing output masks only after verification passes. The novel operations introduced by our protocols---truncation and DP-based polynomial evaluation---are purely local computations that do not introduce new attack surfaces.

\textbf{Theorem 1.}
\textit{Assuming the existence of a PRF, $\pi_{\text{multTrun}}$ (Fig.~\ref{proto:multTrun}) securely realizes the multiplication-with-truncation functionality with computational non-colluding security and fairness in the HA-MSDM model.}

\textbf{Theorem 2.}
\textit{Assuming the existence of a PRF, $\pi_{\text{poly\_fixed}}$ (Fig.~\ref{proto:poly_fixed}) securely realizes the fixed-point polynomial evaluation functionality with computational non-colluding security and fairness in the HA-MSDM model.}

The formal proofs, including explicit simulator constructions for both corruption cases, are provided in Appendix~II of the supplementary material.

\section{Experiments}\label{sec:experiments}

\subsection{Experimental Setup}

\textbf{Implementation.} We implemented our framework in C++20, with the MPC protocol codebase developed following the code style of MD-ML~\cite{yuan2024md}. The training pipeline for polynomial-activation models is implemented in PyTorch, producing exported model weights that are loaded by the C++ inference framework. All experiments were conducted on a server running Ubuntu 20.04.5 LTS, equipped with a 2.2\,GHz AMD Ryzen Threadripper 3970X 32-Core Processor (64 logical cores) and 251\,GB RAM.

\textbf{Network simulation.} We used the Linux Traffic Control (\texttt{tc}) tool to emulate both LAN and WAN environments. The LAN setting uses 1\,ms network delay, 0.01\% packet loss rate, and 10\,Gbps bandwidth. The WAN setting uses 100\,ms network delay, 1\% packet loss rate, and 100\,Mbps bandwidth.

\textbf{Metrics.} Time overhead is defined as the wall-clock duration from when all participating entities (including HP) begin the task until the last entity completes it. Communication overhead is defined as the total size of messages sent by all participating entities (including HP).

\subsection{Microbenchmarks of MPC Layer}

This section evaluates the core building blocks of our MPC layer. We also conducted microbenchmarks on individual CNN building blocks (convolution, ReLU, pooling, BN layers); these results are provided in Appendix~V-B of the supplementary material.

We compare three approaches for evaluating a degree-4 fixed-point polynomial (the ReLU approximation in Equation~(\ref{eq:poly_relu})) on a vector of 65{,}536 elements:
\begin{enumerate}
\item \textbf{Horner's method} using sequential $\pi_{\text{multTrun}}$, requiring $k = 4$ communication rounds;
\item \textbf{Binary tree} method that organizes power computations into a balanced tree with $\lceil \log_2 k \rceil = 2$ rounds;
\item $\boldsymbol{\pi_{\textbf{poly\_fixed}}}$ \textbf{(Ours)} that computes all powers locally via the DP recurrence with 2 fixed communication rounds.
\end{enumerate}

\begin{figure*}[!t]
\centering
\includegraphics[width=\textwidth]{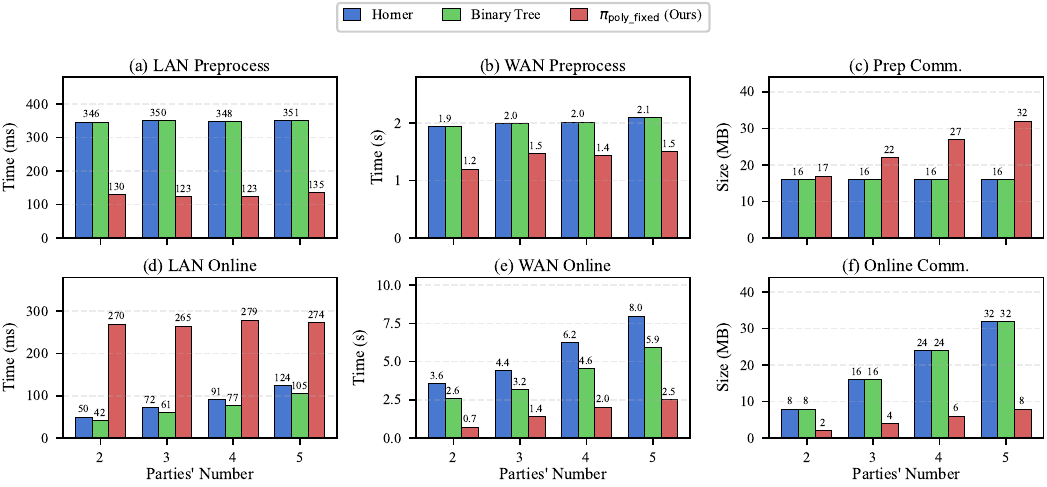}
\caption{Preprocessing and online benchmarks of three fixed-point polynomial evaluation approaches for degree-4 ReLU approximation (65{,}536 elements).}
\label{fig:poly_comparison}
\end{figure*}

Fig.~\ref{fig:poly_comparison} presents the time and communication overhead for 2--5 parties.

In the WAN setting, $\pi_{\text{poly\_fixed}}$ achieves the lowest total time across all party counts, with 2.2--2.8$\times$ speedup over Horner and 1.7--2.4$\times$ over the binary tree method. The advantage comes from both phases: the preprocessing is faster because HP can generate all masks in a single batch, and the online phase benefits from only 2 fixed communication rounds with 4$\times$ lower communication (2--8\,MB vs 8--32\,MB).

In the LAN setting, the total times of all three methods are comparable (396--475\,ms). Although $\pi_{\text{poly\_fixed}}$ has a higher online time due to the $O(k^2)$ local DP computation, this is offset by its faster preprocessing. Since WAN is more representative of real-world deployment, $\pi_{\text{poly\_fixed}}$ is the preferred choice for private inference.

\subsubsection{Accuracy of $\pi_{\text{poly\_fixed}}$}
 
We evaluated the degree-4 ReLU approximation polynomial on 14{,}001 uniformly spaced points over $[-7, 7]$ using a two-party execution and compared the protocol output against the exact polynomial value. As shown in TABLE~\ref{tab:poly_accuracy}, all error metrics remain well below the theoretical per-invocation upper bound of $2^{-d} = 2^{-12} \approx 2.44 \times 10^{-4}$ established in Proposition~3. At representative points including both interval boundaries and the midpoint, the protocol output matches the exact value with zero error, confirming that the truncation carry $\varepsilon \in \{0,1\}$ is the sole error source.
 
\begin{table}[!t]
\caption{Numerical Accuracy of $\pi_{\text{poly\_fixed}}$ on $[-7, 7]$\label{tab:poly_accuracy}}
\centering
\begin{tabular}{l|c}
\hline
\textbf{Metric} & \textbf{Value} \\
\hline
Max absolute error & $3.21 \times 10^{-4}$ \\
Mean absolute error & $9.25 \times 10^{-5}$ \\
RMSE & $1.14 \times 10^{-4}$ \\
Theoretical max (per invocation) & $2^{-12} \approx 2.44 \times 10^{-4}$ \\
\hline
\end{tabular}
\end{table}

\subsection{Benchmarks of CNN Private Inference}

\subsubsection{Efficiency Analysis}

We evaluate the end-to-end online inference time and communication overhead of our framework against two state-of-the-art MSDM frameworks: MD-SONIC~\cite{zhang2025md} and MD-ML~\cite{yuan2024md}. All experiments are conducted in the two-party setting. TABLE~\ref{tab:inference_efficiency} presents the results on LeNet and AlexNet across MNIST, CIFAR-10, Tiny-ImageNet, and ImageNet datasets.

Compared to MD-SONIC, our framework achieves 2.3--5.3$\times$ speedup in LAN and 1.3--3.0$\times$ in WAN. Compared to MD-ML, the speedup is 3.2--6.8$\times$ in LAN and 2.6--5.6$\times$ in WAN. The speedup is most pronounced on larger networks and datasets, reaching 5.3$\times$ (vs MD-SONIC) and 6.8$\times$ (vs MD-ML) in LAN on AlexNet-ImageNet. Regarding communication overhead, our framework consistently achieves an order-of-magnitude reduction over MD-ML across all configurations. Compared to MD-SONIC, our communication is comparable or slightly lower on small networks such as LeNet, but higher on larger networks such as AlexNet. This is because MD-SONIC adopts TinyOT-style Boolean secret sharing~\cite{burra2021high} for nonlinear layers, which yields a compact bit-level representation that scales favorably when nonlinear operations dominate. In contrast, our framework computes all nonlinear operations through polynomial approximation over arithmetic secret sharing, whose per-element ring communication grows more rapidly with the activation tensor size. Despite this trade-off in communication on large networks, our framework achieves substantially lower end-to-end latency in both LAN and WAN settings.

\textbf{On the comparison with MSDM baselines.} A natural concern is whether the observed speedup stems solely from our weaker HA-MSDM assumption. We argue it does not: (i) HA-MSDM is a realistic MLaaS deployment setting where HP is naturally instantiated by the cloud service provider, no stronger than existing trust in the cloud operator; (ii) our framework additionally achieves fairness rather than abort security, orthogonal to efficiency; (iii) simply plugging a helper into MSDM does not yield our fixed-round polynomial evaluation---that requires the co-design of ring-based authenticated sharing with HP-assisted power preprocessing. The speedup thus reflects genuine protocol-level advances rather than an artifact of the threat model.

\begin{table*}[!t]
\caption{Online Inference Time and Communication Overhead on LeNet and AlexNet (Two-Party Setting)\label{tab:inference_efficiency}}
\centering
\begin{minipage}[t]{0.48\textwidth}
\centering
\begin{tabular}{l|l|c|c|c}
\hline
\multicolumn{5}{c}{\textbf{LeNet}} \\
\hline
\textbf{Dataset} & \textbf{Framework} & \textbf{LAN} & \textbf{WAN} & \textbf{Comm.} \\
\hline
\multirow{3}{*}{MNIST} & Ours & 691ms & 1.94s & 0.39MB \\
 & MD-SONIC & 1.68s & 3.52s & 0.48MB \\
 & MD-ML & 2.20s & 5.25s & 19.13MB \\
\hline
\multirow{3}{*}{CIFAR-10} & Ours & 822ms & 2.07s & 0.54MB \\
 & MD-SONIC & 2.29s & 3.83s & 0.67MB \\
 & MD-ML & 3.83s & 6.37s & 26.58MB \\
\hline
\multirow{3}{*}{\makecell[l]{Tiny-\\ImageNet}} & Ours & 3.65s & 6.17s & 2.74MB \\
 & MD-SONIC & 9.83s & 12.59s & 3.37MB \\
 & MD-ML & 12.75s & 24.14s & 133.22MB \\
\hline
\end{tabular}
\end{minipage}%
\hspace{0.02\textwidth}%
\begin{minipage}[t]{0.48\textwidth}
\centering
\begin{tabular}{l|l|c|c|c}
\hline
\multicolumn{5}{c}{\textbf{AlexNet}} \\
\hline
\textbf{Dataset} & \textbf{Framework} & \textbf{LAN} & \textbf{WAN} & \textbf{Comm.} \\
\hline
\multirow{3}{*}{CIFAR-10} & Ours & 3.30s & 8.27s & 21.98MB \\
 & MD-SONIC & 7.58s & 11.14s & 2.30MB \\
 & MD-ML & 11.48s & 21.51s & 103.57MB \\
\hline
\multirow{3}{*}{\makecell[l]{Tiny-\\ImageNet}} & Ours & 46.40s & 88.67s & 93.23MB \\
 & MD-SONIC & 147.69s & 161.01s & 44.55MB \\
 & MD-ML & 197.59s & 314.17s & 1504.84MB \\
\hline
\multirow{3}{*}{ImageNet} & Ours & 61.72s & 113.20s & 70.11MB \\
 & MD-SONIC & 324.29s & 344.37s & 67.87MB \\
 & MD-ML & 417.14s & 633.95s & 2688.20MB \\
\hline
\end{tabular}
\end{minipage}
\end{table*}

\subsubsection{Accuracy Analysis}

TABLE~\ref{tab:accuracy_full} presents the inference accuracy of our framework across five CNN architectures on CIFAR-10, CIFAR-100 and ImageNet, including the full ablation study over four strategy combinations: regularization only (\texttt{none}, equivalent to PILLAR~\cite{diaa2024fast} which employs activation regularization but lacks knowledge distillation and warm starting), with warm start only (\texttt{+ws}), with knowledge distillation only (\texttt{+kd}), and our full strategy (\texttt{+kd+ws}). The ReLU plaintext baseline is listed as the accuracy ceiling.

Three key observations emerge from the results. First, our full strategy (\texttt{+kd+ws}) matches or nearly matches the ReLU baseline across all configurations, with a maximum gap of only 0.5\% (MiniONN cipher) and exact matches in most cases. In contrast, regularization alone leaves a gap of 2.0--5.3\%, confirming the necessity of our proposed techniques. Second, the ablation study reveals that knowledge distillation and warm starting are complementary: on ResNet-18 ImageNet (\texttt{+kd}: +3.6\%, \texttt{+ws}: +3.2\%, combined: +4.2\%), confirming the strategy scales to large-scale datasets. Third, the gap between plaintext and ciphertext inference (Gap~2) is consistently small ($\leq$0.5\%), validating that $\pi_{\text{multTrun}}$ and $\pi_{\text{poly\_fixed}}$ with $l = 88$ introduce negligible computation error.

\begin{table*}[!t]
\caption{Inference Accuracy (\%) with Degree-4 Polynomial Approximation: Ablation Study and Comparison\label{tab:accuracy_full}}
\centering
\begin{tabular}{l|l|cc|cc|cc|cc|cc}
\hline
\multirow{2}{*}{\textbf{Model}} & \multirow{2}{*}{\textbf{Dataset}} & \multicolumn{2}{c|}{\textbf{ReLU}} & \multicolumn{2}{c|}{\textbf{Reg. Only}} & \multicolumn{2}{c|}{\textbf{+WS}} & \multicolumn{2}{c|}{\textbf{+KD}} & \multicolumn{2}{c}{\textbf{+KD+WS (Ours)}} \\
\cline{3-12}
 & & Plain & Cipher & Plain & Cipher & Plain & Cipher & Plain & Cipher & Plain & Cipher \\
\hline
\multirow{2}{*}{ResNet-18} & CIFAR-10 & 94.7 & 94.7 & 92.1 & 92.1 & 92.8 & 92.5 & 94.5 & 94.5 & \textbf{94.7} & \textbf{94.7} \\
 & CIFAR-100 & 75.8 & 75.8 & 70.5 & 70.5 & 72.3 & 72.3 & 75.8 & 75.8 & \textbf{75.8} & \textbf{75.8} \\
\hline
\multirow{2}{*}{ResNet-32} & CIFAR-10 & 94.0 & 94.0 & 90.9 & 90.9 & 90.6 & 90.6 & 93.2 & 91.5 & \textbf{94.0} & \textbf{93.5} \\
 & CIFAR-100 & 69.2 & 69.2 & 65.7 & 65.7 & 68.5 & 68.5 & 68.0 & 66.4 & \textbf{69.2} & \textbf{69.2} \\
\hline
\multirow{2}{*}{VGG-16} & CIFAR-10 & 92.6 & 92.6 & 90.0 & 89.6 & 89.0 & 88.5 & 92.0 & 91.5 & \textbf{92.6} & \textbf{92.6} \\
 & CIFAR-100 & 74.2 & 74.2 & 72.0 & 72.0 & 72.5 & 72.5 & 72.8 & 72.5 & \textbf{74.0} & \textbf{74.0} \\
\hline
ResNet-110 & CIFAR-10 & 93.4 & 93.4 & 91.3 & 91.0 & 93.4 & 93.4 & 92.6 & 92.1 & \textbf{93.4} & \textbf{93.4} \\
\hline
MiniONN & CIFAR-10 & 86.0 & 86.0 & 84.0 & 83.3 & 85.5 & 85.0 & 83.7 & 83.5 & \textbf{86.0} & \textbf{85.5} \\
\hline
ResNet-18 & ImageNet & 69.8 & 69.8 & 65.6 & 65.2 & 68.8 & 67.2 & 69.2 & 68.9 & \textbf{69.8} & \textbf{69.4} \\
\hline
\end{tabular}
\end{table*}
\section{Conclusion}\label{sec:conclusion}

This paper proposes a private inference framework based on the HA-MSDM model. We first design a ring-domain authenticated sharing scheme compatible with HA-MSDM, which preserves MAC soundness under statistical security parameter $s$ while enabling efficient fixed-point arithmetic via bit-shift truncation. We then design three efficient fixed-round MPC protocols: $\pi_{\text{multTrun}}$, $\pi_{\text{poly\_integer}}$, and $\pi_{\text{poly\_fixed}}$. For the inference layer, we propose a co-optimized training strategy that combines activation regularization with knowledge distillation and warm starting to close the accuracy gap between polynomial and ReLU models. Experimental results on seven CNN architectures demonstrate that our framework achieves 2.3--6.8$\times$ speedup in LAN and 1.3--5.6$\times$ speedup in WAN compared to state-of-the-art MSDM frameworks, while matching the ReLU baseline accuracy within 0.5\%.

We outline several potential future research directions: 1) exploring primitive implementations under higher output security guarantees such as guaranteed output delivery, 2) leveraging hardware acceleration such as GPUs to further enhance efficiency and 3) extend the HA-MSDM model to larger networks such as transformers to explore its practicality.

\bibliographystyle{IEEEtran}
\bibliography{ref}

@misc{ref1,
  howpublished = {\url{https://www.solutelabs.com/blog/best-mlaas-platforms}},
  year         = {2025}
}

@inproceedings{ng2023sok,
  title={SoK: Cryptographic neural-network computation},
  author={Ng, Lucien KL and Chow, Sherman SM},
  booktitle={2023 IEEE Symposium on Security and Privacy (SP)},
  pages={497--514},
  year={2023},
  organization={IEEE}
}

@inproceedings{yao1982protocols,
  title={Protocols for secure computations},
  author={Yao, Andrew C},
  booktitle={23rd annual symposium on foundations of computer science (sfcs 1982)},
  pages={160--164},
  year={1982},
  organization={IEEE}
}

@inproceedings{cramer2018spd,
  title={SPD: efficient MPC mod for dishonest majority},
  author={Cramer, Ronald and Damg{\aa}rd, Ivan and Escudero, Daniel and Scholl, Peter and Xing, Chaoping},
  booktitle={Annual International Cryptology Conference},
  pages={769--798},
  year={2018},
  organization={Springer}
}

@inproceedings{yuan2024md,
  title={$\{$MD-ML$\}$: Super fast $\{$Privacy-Preserving$\}$ machine learning for malicious security with a dishonest majority},
  author={Yuan, Boshi and Yang, Shixuan and Zhang, Yongxiang and Ding, Ning and Gu, Dawu and Sun, Shi-Feng},
  booktitle={33rd USENIX Security Symposium (USENIX Security 24)},
  pages={2227--2244},
  year={2024}
}

@article{zhang2025md,
  title={MD-SONIC: Maliciously-Secure Outsourcing Neural Network Inference With Reduced Online Communication},
  author={Zhang, Yansong and Chen, Xiaojun and Dong, Ye and Zhang, Qinghui and Hou, Rui and Liu, Qiang and Chen, Xudong},
  journal={IEEE Transactions on Information Forensics and Security},
  year={2025},
  publisher={IEEE}
}

@inproceedings{koti2021swift,
  title={$\{$SWIFT$\}$: Super-fast and robust $\{$Privacy-Preserving$\}$ machine learning},
  author={Koti, Nishat and Pancholi, Mahak and Patra, Arpita and Suresh, Ajith},
  booktitle={30th USENIX security symposium (USENIX security 21)},
  pages={2651--2668},
  year={2021}
}

@inproceedings{karmakar2024asterisk,
  title={Asterisk: Super-fast MPC with a Friend},
  author={Karmakar, Banashri and Koti, Nishat and Patra, Arpita and Patranabis, Sikhar and Paul, Protik and Ravi, Divya},
  booktitle={2024 IEEE Symposium on Security and Privacy (SP)},
  pages={542--560},
  year={2024},
  organization={IEEE}
}

@inproceedings{cleve1986limits,
  title={Limits on the security of coin flips when half the processors are faulty},
  author={Cleve, Richard},
  booktitle={Proceedings of the eighteenth annual ACM symposium on Theory of computing},
  pages={364--369},
  year={1986}
}

@article{garimella2021sisyphus,
  title={Sisyphus: A cautionary tale of using low-degree polynomial activations in privacy-preserving deep learning},
  author={Garimella, Karthik and Jha, Nandan Kumar and Reagen, Brandon},
  journal={arXiv preprint arXiv:2107.12342},
  year={2021}
}

@inproceedings{diaa2024fast,
  title={Fast and private inference of deep neural networks by co-designing activation functions},
  author={Diaa, Abdulrahman and Fenaux, Lucas and Humphries, Thomas and Dietz, Marian and Ebrahimianghazani, Faezeh and Kacsmar, Bailey and Li, Xinda and Lukas, Nils and Mahdavi, Rasoul Akhavan and Oya, Simon and others},
  booktitle={33rd USENIX Security Symposium (USENIX Security 24)},
  pages={2191--2208},
  year={2024}
}

@inproceedings{lu2019honeybadgermpc,
  title={Honeybadgermpc and asynchromix: Practical asynchronous mpc and its application to anonymous communication},
  author={Lu, Donghang and Yurek, Thomas and Kulshreshtha, Samarth and Govind, Rahul and Kate, Aniket and Miller, Andrew},
  booktitle={Proceedings of the 2019 ACM SIGSAC Conference on Computer and Communications Security},
  pages={887--903},
  year={2019}
}

@inproceedings{mohassel2017secureml,
  title={Secureml: A system for scalable privacy-preserving machine learning},
  author={Mohassel, Payman and Zhang, Yupeng},
  booktitle={2017 IEEE symposium on security and privacy (SP)},
  pages={19--38},
  year={2017},
  organization={IEEE}
}

@inproceedings{damgaard2012multiparty,
  title={Multiparty computation from somewhat homomorphic encryption},
  author={Damg{\aa}rd, Ivan and Pastro, Valerio and Smart, Nigel and Zakarias, Sarah},
  booktitle={Annual cryptology conference},
  pages={643--662},
  year={2012},
  organization={Springer}
}

@inproceedings{damgaard2013practical,
  title={Practical covertly secure MPC for dishonest majority--or: breaking the SPDZ limits},
  author={Damg{\aa}rd, Ivan and Keller, Marcel and Larraia, Enrique and Pastro, Valerio and Scholl, Peter and Smart, Nigel P},
  booktitle={European Symposium on Research in Computer Security},
  pages={1--18},
  year={2013},
  organization={Springer}
}

@inproceedings{keller2016mascot,
  title={MASCOT: faster malicious arithmetic secure computation with oblivious transfer},
  author={Keller, Marcel and Orsini, Emmanuela and Scholl, Peter},
  booktitle={Proceedings of the 2016 ACM SIGSAC Conference on Computer and Communications Security},
  pages={830--842},
  year={2016}
}

@inproceedings{keller2018overdrive,
  title={Overdrive: Making SPDZ great again},
  author={Keller, Marcel and Pastro, Valerio and Rotaru, Dragos},
  booktitle={Annual International Conference on the Theory and Applications of Cryptographic Techniques},
  pages={158--189},
  year={2018},
  organization={Springer}
}

@inproceedings{orsini2020overdrive2k,
  title={Overdrive2k: Efficient secure MPC over from somewhat homomorphic encryption},
  author={Orsini, Emmanuela and Smart, Nigel P and Vercauteren, Frederik},
  booktitle={Cryptographers’ track at the RSA conference},
  pages={254--283},
  year={2020},
  organization={Springer}
}

@inproceedings{escudero2022more,
  title={More efficient dishonest majority secure computation over Z 2 k via galois rings},
  author={Escudero, Daniel and Xing, Chaoping and Yuan, Chen},
  booktitle={Annual International Cryptology Conference},
  pages={383--412},
  year={2022},
  organization={Springer}
}

@inproceedings{chandran2019universally,
  title={Universally composable secure computation with corrupted tokens},
  author={Chandran, Nishanth and Chongchitmate, Wutichai and Ostrovsky, Rafail and Visconti, Ivan},
  booktitle={Annual International Cryptology Conference},
  pages={432--461},
  year={2019},
  organization={Springer}
}

@inproceedings{hazay2016composable,
  title={Composable security in the tamper-proof hardware model under minimal complexity},
  author={Hazay, Carmit and Polychroniadou, Antigoni and Venkitasubramaniam, Muthuramakrishnan},
  booktitle={Theory of Cryptography Conference},
  pages={367--399},
  year={2016},
  organization={Springer}
}

@inproceedings{badrinarayanan2018non,
  title={Non-interactive secure computation from one-way functions},
  author={Badrinarayanan, Saikrishna and Jain, Abhishek and Ostrovsky, Rafail and Visconti, Ivan},
  booktitle={International Conference on the Theory and Application of Cryptology and Information Security},
  pages={118--138},
  year={2018},
  organization={Springer}
}

@inproceedings{ishai2022fully,
  title={Fully-secure MPC with minimal trust},
  author={Ishai, Yuval and Patra, Arpita and Patranabis, Sikhar and Ravi, Divya and Srinivasan, Akshayaram},
  booktitle={Theory of Cryptography Conference},
  pages={470--501},
  year={2022},
  organization={Springer}
}

@article{muth2022assisted,
  title={Assisted mpc},
  author={Muth, Philipp and Katzenbeisser, Stefan},
  journal={Cryptology ePrint Archive},
  year={2022}
}

@inproceedings{damgaard2019new,
  title={New primitives for actively-secure MPC over rings with applications to private machine learning},
  author={Damg{\aa}rd, Ivan and Escudero, Daniel and Frederiksen, Tore and Keller, Marcel and Scholl, Peter and Volgushev, Nikolaj},
  booktitle={2019 IEEE Symposium on Security and Privacy (SP)},
  pages={1102--1120},
  year={2019},
  organization={IEEE}
}

@inproceedings{gilad2016cryptonets,
  title={Cryptonets: Applying neural networks to encrypted data with high throughput and accuracy},
  author={Gilad-Bachrach, Ran and Dowlin, Nathan and Laine, Kim and Lauter, Kristin and Naehrig, Michael and Wernsing, John},
  booktitle={International conference on machine learning},
  pages={201--210},
  year={2016},
  organization={PMLR}
}

@article{chabanne2017privacy,
  title={Privacy-preserving classification on deep neural network},
  author={Chabanne, Herv{\'e} and De Wargny, Amaury and Milgram, Jonathan and Morel, Constance and Prouff, Emmanuel},
  journal={Cryptology ePrint Archive},
  year={2017}
}

@article{hesamifard2018privacy,
  title={Privacy-preserving machine learning as a service},
  author={Hesamifard, Ehsan and Takabi, Hassan and Ghasemi, Mehdi and Wright, Rebecca N},
  journal={Proceedings on Privacy Enhancing Technologies},
  year={2018}
}

@inproceedings{mishra2020delphi,
  title={Delphi: A cryptographic inference system for neural networks},
  author={Mishra, Pratyush and Lehmkuhl, Ryan and Srinivasan, Akshayaram and Zheng, Wenting and Popa, Raluca Ada},
  booktitle={Proceedings of the 2020 workshop on privacy-preserving machine learning in practice},
  pages={27--30},
  year={2020}
}

@article{lee2023precise,
  title={Precise approximation of convolutional neural networks for homomorphically encrypted data},
  author={Lee, Junghyun and Lee, Eunsang and Lee, Joon-Woo and Kim, Yongjune and Kim, Young-Sik and No, Jong-Seon},
  journal={IEEE Access},
  volume={11},
  pages={62062--62076},
  year={2023},
  publisher={IEEE}
}

@article{knott2021crypten,
  title={Crypten: Secure multi-party computation meets machine learning},
  author={Knott, Brian and Venkataraman, Shobha and Hannun, Awni and Sengupta, Shubho and Ibrahim, Mark and van der Maaten, Laurens},
  journal={Advances in Neural Information Processing Systems},
  volume={34},
  pages={4961--4973},
  year={2021}
}

@inproceedings{patra2021aby2,
  title={$\{$ABY2. 0$\}$: Improved $\{$mixed-protocol$\}$ secure $\{$two-party$\}$ computation},
  author={Patra, Arpita and Schneider, Thomas and Suresh, Ajith and Yalame, Hossein},
  booktitle={30th USENIX Security Symposium (USENIX Security 21)},
  pages={2165--2182},
  year={2021}
}

@article{hinton2015distilling,
  title={Distilling the Knowledge in a Neural Network},
  author={Hinton, Geoffrey and Vinyals, Oriol and Dean, Jeff},
  journal={stat},
  volume={1050},
  pages={9},
  year={2015}
}

@article{burra2021high,
  title={High-performance multi-party computation for binary circuits based on oblivious transfer},
  author={Burra, Sai Sheshank and Larraia, Enrique and Nielsen, Jesper Buus and Nordholt, Peter Sebastian and Orlandi, Claudio and Orsini, Emmanuela and Scholl, Peter and Smart, Nigel P},
  journal={Journal of Cryptology},
  volume={34},
  number={3},
  pages={34},
  year={2021},
  publisher={Springer}
}

@inproceedings{damgaard2007scalable,
  title={Scalable and unconditionally secure multiparty computation},
  author={Damg{\aa}rd, Ivan and Nielsen, Jesper Buus},
  booktitle={Annual International Cryptology Conference},
  pages={572--590},
  year={2007},
  organization={Springer}
}

@inproceedings{damgaard2006unconditionally,
  title={Unconditionally secure constant-rounds multi-party computation for equality, comparison, bits and exponentiation},
  author={Damg{\aa}rd, Ivan and Fitzi, Matthias and Kiltz, Eike and Nielsen, Jesper Buus and Toft, Tomas},
  booktitle={Theory of Cryptography Conference},
  pages={285--304},
  year={2006},
  organization={Springer}
}

@inproceedings{ben2019turbospeedz,
  title={Turbospeedz: Double your online SPDZ! Improving SPDZ using function dependent preprocessing},
  author={Ben-Efraim, Aner and Nielsen, Michael and Omri, Eran},
  booktitle={International Conference on Applied Cryptography and Network Security},
  pages={530--549},
  year={2019},
  organization={Springer}
}

@article{harth2025sok,
  title={SoK: Truncation Untangled: Scaling Fixed-Point Arithmetic for Privacy-Preserving Machine Learning to Large Models and Datasets},
  author={Harth-Kitzerow, Christopher and Suresh, Ajith and Carle, Georg},
  journal={Proceedings on Privacy Enhancing Technologies},
  year={2025}
}

\end{document}